\documentclass[journal]{IEEEtran}
\IEEEoverridecommandlockouts
\usepackage[T1]{fontenc}% optional T1 font encoding
\usepackage[noadjust]{cite}
\usepackage{amsmath,amssymb,amsfonts}
\usepackage{algorithm}
\usepackage{algorithmic}
\usepackage{amsthm}
\usepackage{hyperref}
\theoremstyle{definition}
\newtheorem{definition}{Definition}
\theoremstyle{plain}

\usepackage{graphicx}
\usepackage{textcomp}
\usepackage{xcolor}
\usepackage[inline]{enumitem}
\usepackage{float}
\newfloat{algorithm}{t}{lop}

\usepackage[flushleft]{threeparttable}
\usepackage{array,booktabs,makecell}
\usepackage[inline]{enumitem}

\usepackage{orcidlink}
\usepackage{siunitx}

\def\BibTeX{{\rm B\kern-.05em{\sc i\kern-.025em b}\kern-.08em
		T\kern-.1667em\lower.7ex\hbox{E}\kern-.125emX}}

\ifCLASSINFOpdf
\else
\fi
\usepackage{amsmath}
\usepackage{caption}
\usepackage{subcaption}

\begin{document}
\title{UAV-Aided Lifelong Learning for AoI and Energy Optimization in Non-Stationary IoT Networks}
% \author{XXX, XXX, and XXX %\IEEEmembership{Member,~IEEE,}
\author{Zhenzhen~Gong\orcidlink{https://orcid.org/0009-0003-5525-0410},
	   Omar~Hashash\orcidlink{https://orcid.org/0000-0003-0086-5250},~\IEEEmembership{Graduate~Student~Member,~IEEE,}
        Yingze~Wang\orcidlink{https://orcid.org/0000-0002-5966-7139},
	 Qimei~Cui\orcidlink{https://orcid.org/0000-0003-1720-220X},~\IEEEmembership{Senior~Member,~IEEE,}
	   Wei~Ni\orcidlink{https://orcid.org/0000-0002-4933-594X},~\IEEEmembership{Senior~Member,~IEEE,}
     Walid~Saad\orcidlink{https://orcid.org/0000-0003-2247-2458},~\IEEEmembership{Fellow,~IEEE,}
	  and~Kei~Sakaguchi\orcidlink{https://orcid.org/0000-0002-9334-2477},~\IEEEmembership{Senior~Member,~IEEE \vspace{-0.8cm}} 
       % <-this % stops a space
\thanks{Zhenzhen Gong, Yingze Wang and Qimei Cui are with the National Engineering Laboratory for Mobile Network Technologies, Beijing University of Posts and Telecommunications, Beijing 100876, China 
(email: \{gzz0822, wang\_ying\_ze1993, cuiqimei\}@bupt.edu.cn).}
% e-mail: (see http://www.michaelshell.org/contact.html).}% <-this % stops a space
\thanks{Omar Hashash and Walid Saad are with Wireless@VT, Bradley Department of Electrical and Computer Engineering, Virginia Tech, Arlington, VA, 22203, USA (email: \{omarnh, walids\}@vt.edu).}% <-this % stops a space
% \thanks{Manuscript received April 19, 2005; revised August 26, 2015.}
\thanks{Wei Ni is with Data61, CSIRO, Marsfield, NSW 2122, Australia (email: Wei.Ni@data61.csiro.au)}
\thanks{Kei Sakaguchi is with Department of Electrical and Electronic Engineering, Tokyo Institute of Technology, Tokyo, 152-8550, Japan (email: sakaguchi@mobile.ee.titech.ac.jp).}
}

\maketitle

\begin{abstract}
	In this paper, a novel joint energy and age of information (AoI) optimization framework for IoT devices in a \emph{non-stationary} environment is presented. In particular, IoT devices that are distributed in the real-world are required to efficiently utilize their computing resources so as to balance the freshness of their data and their energy consumption. To optimize the performance of IoT devices in such a dynamic setting, a novel \emph{lifelong reinforcement learning (RL)} solution that enables IoT devices to continuously adapt their policies to each newly encountered environment is proposed. Given that IoT devices have limited energy and computing resources, an unmanned aerial vehicle (UAV) is leveraged to visit the IoT devices and update the policy of each device sequentially. As such, the UAV is exploited as a mobile learning agent that can learn a shared knowledge base with a feature base in its training phase, and feature sets of a zero-shot learning method in its testing phase, to generalize between the environments. To optimize the trajectory and flying velocity of the UAV, an actor-critic network is leveraged so as to minimize the UAV energy consumption. Simulation results show that the proposed lifelong RL solution can outperform the state-of-art benchmarks by enhancing the balanced cost of IoT devices by $8.3\%$ when incorporating warm-start policies for unseen environments. In addition, our solution achieves up to $49.38\%$ reduction in terms of energy consumption by the UAV in comparison to the random flying strategy. \vspace{-0.1cm}
\end{abstract}

% To do in abstract:
%  simulation: benchmarks if possible
%  clear about the performance improvement: Here it is a little bit unclear. How does it outperform? Like in terms of energy, latency, the reward? state what 60% and 80% refer to here.

\begin{IEEEkeywords}
Internet of Things (IoT), Unmanned Aerial Vehicle (UAV), Age of Information (AoI), Lifelong Learning. 
%Reinforcement Learning (RL), Actor-Critic (AC)
\vspace{-0.38cm}
\end{IEEEkeywords}

% \fi
\IEEEpeerreviewmaketitle

\section{Introduction}
\vspace{-0.1cm}

\IEEEPARstart{T}{he} Internet of Things (IoT) \cite{gong2021lifelong} represents a technological breakthrough that brings forth numerous opportunities for new applications at the intersection of wireless communications and intelligent industries, e.g., Industry 4.0 \cite{chaccour2022seven}. 
In fact, IoT devices can provide increased levels of autonomy to physical systems by harnessing their immense capabilities to sense and collect data from their surroundings \cite{hashash2022edge}.
% Harnessing the capability of sensing and collecting data from their surroundings, IoT devices are able to provide increased autonomy to physical systems and interoperability to our daily life. 
In essence, capturing the performance of IoT devices in such use cases has been an active area of research and recent interest \cite{stoyanova2020survey}. Essentially, this requires adopting novel metrics such as the \emph{age of information (AoI)} to reflect the timeliness and freshness of the underlying physical systems being monitored \cite{kaul2012real}.
% Here the AoI is defined as the time elapsed since the data packages have been generated from the physical source.
Here, the AoI is defined as the time elapsed since the last successfully received update packet at the IoT device was generated by the physical source \cite{abd2019role}.
Nevertheless, IoT devices often lack sufficient computing capabilities and have limited energy resources to offload their collected data to remote base stations (BSs). Hence, it is challenging for IoT devices to operate in areas with poor connectivity and provide reliable services for mission-critical physical systems. 
To alleviate such challenges, one can integrate IoT networks with unmanned aerial vehicles (UAVs) \cite{alzahrani2020uav, islam2021review, amrallah2023uav} to improve wireless connectivity and enhance computing abilities.
% Furthermore, unmaned aerial vehicles (UAVs), also known as drones, 
Thus, UAVs can be deployed as flying BSs that can communicate with IoT devices in a cost-efficient way. 
% UAVs have gained popularity for their flexibility and potential use in various applications\cite{mozaffari2019tutorial}, such as earthquake rescue, property surveillance, and emergency communication \cite{alzahrani2020uav, muchiri2022review}. 
Hence, UAVs assist IoT networks in many applications
%  \cite{mozaffari2019tutorial} 
such as rescue services \cite{alzahrani2020uav} and smart farming \cite{islam2021review}.
% can also enhance the intelligence of future applications, such as Smart City, Internet of Things (IoT), and 6G, by providing extra close-to-site view and dynamic tracking.
% To overcome this challenge, UAVs can be deployed as flying BSs that can communicate with IoT devices in a cost-efficient way. 
Evidently, UAVs can enhance the performance of IoT networks
% and save energy for IoT devices 
by providing versatile wireless and computing services to aid autonomous decision making
% (e.g., through reinforcement learning (RL)) \cite{mozaffari2018beyond}, 
at the IoT level, and simultaneously, reduce the energy cost of IoT devices.
% \cite{mozaffari2018beyond, chen2019artificial, chen2017caching}

Despite the surge on IoT in literature \cite{stoyanova2020survey, hu2020aoi, laghari2021review}remarkably,
% {zhou2019joint, hu2020aoi, 
% However, remarkably, many prior works \cite{ditzler2015learning, parisi2019continual, bommasani2021opportunities}, often assume 
the physical environment associated with IoT networks remains to be considered stationary -- \emph{an ideal assumption that rarely holds true in practice} \cite{thrun1998lifelong}.
% , which rarely holds true in practice \cite{thrun1998lifelong}. 
As such, this assumption normally considers that the generation of data from the physical system follows a fixed distribution \cite{ditzler2015learning}
% , parisi2019continual, bommasani2021opportunities}. 
Consequently, the optimal strategies and policies that govern the operation of IoT devices are  then considered to be time-invariant~\cite{padakandla2021survey}.
%  and would not change over time~\cite{padakandla2021survey}. 
On the contrary, due to changes in the 
% users’ habits or preferences, geographical 
environment affecting the physical system (e.g., thermal drifts) or on the system level itself (e.g., aging effects), \emph{non-stationary} conditions for IoT devices arise accordingly \cite{o2014anomaly}. 
%  even thermal drifts or aging effects in IoT devices.
Thus, it remains
% In practice, geographical environments and weather conditions change over time, making it 
challenging for IoT devices to optimize their performance and reduce their AoI and energy costs in such real-world scenarios~\cite{padakandla2020reinforcement}.
% , shafi2019precision, kober2013reinforcement, niroui2019deep}. 
% Let alone dynamic environments that are inherent non-stationary \cite{guan2021uav}. 
%% A specialized behavior strategy is required for each different environment, which can significantly impact the performance and effectiveness of UAVs in IoT applications.
% from: https://arxiv.org/pdf/2108.07258.pdf
%搜索stationary的部分：
% to keep a model’s knowledge continually up-to-date with world events or cultural developments, continually add data from completely new domains or modalities as they become available, or continually edit a model’s memories to comply with privacy or legal constraints as a society’s values or laws evolve
Noticeably, relying on conventional methods such as reinforcement learning (RL) to do so by optimizing the operating policies of these devices drastically fails to address the challenges posed by non-stationary environments. This stems from the fact that such solutions are theoretically developed and tailored towards operating in stationary environments, whereby any variabilities in the environment can lead to suboptimal performance and degradation in the system reliability levels. 
Henceforth, a robust RL solution that can \emph{continuously adapt the policies} of IoT devices to unprecedented \emph{non-stationary} developments in the environment is ultimately required.

\vspace{-0.4cm}
\subsection{Prior Works}

Minimizing the AoI has been extensively addressed in various IoT network scenarios \cite{zhou2019joint, wang2020minimizing, hatami2020age} with the aid of UAVs \cite{ferdowsi2020neural, zhou2019deep}. In particular, the works in \cite{ferdowsi2020neural} and \cite{zhou2019deep} investigate minimizing the AoI for IoT scheduling updates, while optimizing the trajectory \cite{ferdowsi2020neural} and velocity \cite{zhou2019deep} of the UAV.
% energy harvesting systems \cite{hatami2020age}, vehicular networks \cite{kaul2011minimizing}, cognitive networks \cite{wang2020minimizing} and augmented reality services \cite{chaccour2020ruin}.
Nevertheless, these works \cite{ ferdowsi2020neural, wang2020minimizing, hatami2020age} rely on classical RL approaches (e.g., Q-learning) throughout their solutions. Accordingly, this limits the novelty of these solutions to ideal stationary scenarios. Furthermore, such stationary assumption also affects the practical UAV functionality in assisting IoT networks. In practice, leveraging RL solutions (e.g., value-based learning \cite{sutton2018reinforcement} and policy-based learning \cite{goodfellow2016deep}) to optimize the UAV operations and trajectory in such non-stationary scenarios can lead to unstable rewards while draining computing resources.
To address this issue, recent works such as \cite{xie2020deep} and \cite{abdallah2016addressing}, have discussed RL solutions for a broad range of applications in a non-stationary setting.
Nevertheless, the works in \cite{xie2020deep} and \cite{abdallah2016addressing} have key limitations that hinder their practicality.
In fact, these works assume that the evolution of non-stationary environments is predictable in nature. Thus, these works assume full knowledge about the period of each environment encountered and its transitions.
For instance, the work in \cite{xie2020deep} models the non-stationary environments as continually evolving Markov decision processes (MDPs) that are based on a soft actor-critic (AC) architecture. 
While such approach \cite{xie2020deep} works explicitly for \emph{periodically} changing non-stationary environments, it is clearly not suitable for capturing the \emph{abrupt, random, and miscellaneous} transitions that occur in dynamic real-world environments.
Moreover, the variants of Q-learning emerging in \cite{abdallah2016addressing} have shown inefficient performance on multiple fronts.
On the one hand, the previously learned parameters and optimal policies are \emph{discarded} for each new environment encountered. On the other hand, adapting the policies can be computationally intense and resource draining until \emph{convergence} is reached.
Furthermore, merging RL with other fundamental methods that provide a leap into the non-stationary realm has also been limited. Here, one of these prominent solutions can be to embrace RL with meta-learning \cite{finn2017model} and transfer learning. In particular, this solution can considerably maintain adequate performance in new environments with the aid of a few gradient updates and $k$-shot learning. 
% However, a major drawback of this solution lies in its long convergence time when the new environments are relatively non-similar and statistically apart from each other. 
%Indeed, maintaining the acquired knwowledge will harness additional benefits for such real-time applications.
Nonetheless, it tends to \emph{lose the knowledge} acquired about previously encountered environments with each new update. In fact, harnessing this knowledge can be a key to enable swift updates \cite{kumar2012learning}. Indeed, exploiting other methods that fill in this gap and facilitate maintaining continuous knowledge transfer (e.g., through a knowledge base \cite{yang2010image}) between environments in a RL regime can provide substantial enhancements \cite{xie2020deep, abdallah2016addressing, finn2017model}. However, they come at the expense of \emph{changing the architecture} of the knowledge base for each new environment encountered, which can be largely inefficient.

In essence, an effective RL solution in a non-stationary setting demands a rigorous design \emph{to detect transitions between environments, accumulate the acquired knowledge and successfully transfer it between environments, while ensuring  a minimal convergence time for updates and the sustainability of acquired parameters and architectures.} To this end, this novel RL solution must generalize between environments on the fly, while efficiently utilizing computing and energy resources of both the UAV and IoT devices, simultaneously. Hence, such a solution should consider the optimization of the UAV inherently within its design, as the non-stationary dimension in the IoT network will impinge on the operating strategy and functionality of the UAV. As such, fulfilling these requirements demands a shift towards \emph{lifelong learning}~\cite{chen2018lifelong} that can provide a \emph{continuously evolving, knowledge-aware, and generalizable} RL solution in a non-stationary setting.
% As such, fulfilling these requirements falls along the lines of lifelong learning that is able to provide a continuous and efficient method for non-stationary IoT networks.

\vspace{-0.3cm}
\subsection{Contributions} %Omar Version

The main contribution of this paper is the development of a novel UAV-aided \emph{lifelong RL} approach to continuously optimize the data freshness and energy efficiency of IoT devices in non-stationary environments. In particular, IoT devices distributed in the real-world must effectively utilize their computing resources to balance their cost of AoI and energy consumption. This is carried out in the presence of a UAV that facilitates the rapid adaptation of these IoT devices to their dynamic environments through updating their operating policies accordingly. Here, the UAV acts as a central mobile agent responsible for visiting each IoT device while updating and learning a knowledge basis and feature basis of the encountered environments, respectively. Thus, our proposed lifelong approach further focuses on developing a generalizable model that can seamlessly adapt to new environments without altering the underlying knowledge base. In fact, such model with mapping vectors for each environment individually enables the separation of the shared knowledge and environment specific knowledge. Hence, this enables the rapid adaptation to new environments by alleviating the need for extensive modifications or retraining of the model. To efficiently utilize the energy resources of the UAV in this operation, an AC network is leveraged to optimize its flying trajectory and velocity between the devices.
By utilizing the accumulated knowledge basis and the extracted features, the framework can effectively determine the best IoT interacting policies for different environments without the need for explicit training on every specific environment.
%Furthermore, our approach emphasizes the non-stationarity of natural environments, which is inherently unpredictable. In the proposed approach, the UAV's access is limited to the device it is currently visiting, making it impractical to establish a global agent that has complete knowledge of all devices simultaneously. This approach aligns more closely with the practical constraints of real-world scenarios.
\emph{To the best of our knowledge, this is the first work that considers the joint optimization of both IoT devices and UAV in non-stationary environments.}

In summary, our key contributions include:
\begin{itemize}
	\item We propose a UAV assisted IoT framework that enables the UAV to help IoT devices adjust to non-stationary environments while minimizing its flying energy. In this framework, the energy consumption of the UAV and the cost of the IoT devices (represented in terms of AoI and computing energy) are jointly considered. A fix and optimize method is applied for better analysis. 
	\item We propose a novel lifelong RL approach that enables knowledge transfer for a stream of environments. By employing this approach, a small amount of sampled data can facilitate convergence within a few steps. In addition, the use of zero-shot method enables the quick feature extraction such that a warm-start initial policy can be provided for unseen environments.
	% when combined with a warm-start policy as the initial policy. 
    \item A novel environment discovery method is explored to detect the environment change point. Using the sampled information, the application of feature extraction and a doubled knowledge basis together can enable fast knowledge extraction for unseen environments.  
\end{itemize}

Simulation results showcase that our proposed lifelong RL solution can reduce up to $50\%$ of the convergence time for updates in comparison to random initial policies. Moreover, it can reduce the energy consumption of the UAV by $49.38\%$ in comparison to the random flying strategy.
Furthermore, in our previous work~\cite{gong2021lifelong}, we have developed a lifelong RL solution to optimize of the AoI and computing energy of IoT devices in a non-stationary setting. However, this early work did not consider the influence of the non-stationary environment on the UAV flying strategy. In contrast, this work comprehensively addresses the flight strategy of the UAV and inherently incoporates it within our lifelong RL approach.
% In contrast, this journal version provides a comprehensive examination of the UAV's flight strategies and their integration with our lifelong RL approach.

The rest of the paper is organized as follows.
The system model and problem formulation are provided in Section~\ref{sec:Sys mdl}.
The lifelong RL algorithm to optimize IoT devices in non-stationary environments is presented in Section~\ref{sec:LL slution}.
The design of an energy efficient solution to optimize the trajectory and velocity of the UAV is presented in Section~\ref{sec: UAV energy}.
Simulation results are provided in Section~\ref{sec:simulation}. Finally, conclusions are drawn in Section~\ref{sec:conclusion}.
The notations used in this paper are shown in Table \ref{Table1}.

\begin{table}
	\caption{Parameters and notations of the system model.}
	\centering 
	\begin{threeparttable}
	\begin{tabular}{lll}
	%{m{15mm} m{70mm} m{18mm}}
	Parameters  & Descriptions  & \\
	\midrule\midrule $N$  &  Number of IoT Devices &   \\
	$\mathcal{N}$ & The set of IoT Devices &  \\ 
	i & The index of device &  \\ 
	$t$ &  The index of time slot &  \\
	$t^s_{i,t}$, $t^e_{i,t}$ & Starting and ending time of the environment \\
	$\boldsymbol{d}_i$ &   The Cartesian coordinates of device $i$    \\
	$\lambda_{i,t}$ & Probability of the  Bernoulli distribution & \\
	$q_{i,t}$ & The indicator of the package arrival \\
	$a_{i,t}$ &  The size of the data packet &   \\
	$\bar{a}_{i,t}$ &   Mean value of data package size  &  \\
	$\sigma_{i,t}$ &  Standard deviation of Gaussian distribution  &  \\
	$\epsilon_{i,t}$ &  Number of CPU cycles for device $i$ at time slot $t$  &  \\
	$\epsilon_{i,\textrm{max}}$  &  Maximum number of CPU cycles for device $i$  &  \\
	$\kappa_i$ & Device's chip architecture related parameter  &  \\
	$b_{i,t}$ &  Queues of unprocessed packages in CPU cycles  &  \\
	$\Delta_{i, t}$ &  AoI of device $i$ at time slot $t$ &  \\
	$\xi_{i,t}$ & Index of the most recently processed data packet &  \\
	$u_{\xi_{i,t}}$ &  The generation time of packet $\xi_{i,t}$ &  \\
	$\omega_{i,t+1}$ &  Indicator of the empty queue for time slot $t$ &  \\
	$c_i(t)$ & Cost of device $i$ at time slot $t$ &  \\
	$\beta, \mu, \eta_x$ & Trade-off factors with $x = {1, 2, 3, 4}$ \\
	$\emph{z}_{i,t}$ &  Environment specific set of device $i$ at slot $t$  &  \\
	$\boldsymbol{l}_0$ & UAV's initial location &  \\
	$\boldsymbol{l}_m$ & UAV's location in its $m-$th decision &  \\
	$v_m$ &  Velocity of the UAV for its $m-$th flight &  \\
	$v_{\textrm{min}}, v_{\textrm{max}}$ &  UAV's minimum and maximum velocity &  \\
	$\boldsymbol{\varepsilon}_i$ & The CPU decision vector of device $i$\\
	$\boldsymbol{E}$ & The CPU decision matrix of all the devices \\
	$\boldsymbol{v}$ & The vector of UAV's velocity during flight \\
	$\boldsymbol{F}$ & The vector of UAV's destination during flight \\
	$\epsilon_z$ & Gaussian noise of stochastic policy\\
	$\sigma_z$ & Standard deviation of Gaussian distribution \\
	$e_0$ &  Propulsion energy per unit time  &  \\
	$e^U_m$ &  Total energy consumption
	of UAV's $m-$th  flight  &  \\
	$\boldsymbol{\tau}$ & Interaction history collected from environment\\
	$T$ &  The total number of time slots &  \\
	$n_i$ & Number of environments of device $i$ &  \\
	$Z$ & Number of environments for all the devices &  \\
	$M$ & Number of flying decisions of the UAV &  \\
	\midrule\midrule
	\end{tabular}\vspace{-0.4cm}
	% \begin{tablenotes}
	% \item[*] Throughout the paper. 
	% \end{tablenotes}
	\end{threeparttable}
	\label{Table1}  
\end{table}

% \begin{table}[H]
%    \caption{Example LaTeX Table} 
%    \label{tab:example}
%    \small
%    \centering
%    \begin{tabular}{l|c|c|r}
%    \toprule
%    \textbf{Header 1} & \textbf{Header 2} & \textbf{Header 3} & \textbf{Header 4} \\ 
%    \midrule
%    Item 1 & Item 2 & Item 3 & Item 4 \\
%    \midrule
%    Item 1 & Item 2 & Item 3 & Item 4\\
%    \bottomrule
%    \end{tabular}
% \end{table}

% \begin{figure}[htbp]
% 	\centerline
% 	{\includegraphics[width=0.5\textwidth]{General_DT-Models.pdf}}
% 	\caption{Key design requirements of the UAV based intelligent DT.}
% 	\label{General_DT}
% \end{figure}

% With the advantages mentioned above, a DT enabled intelligent adaptation framework is designed to facilitate the non-stationary UAV-Devices system. 
% In our system, the UAV can be an intelligent center that is equipped with a virtual twin space. As such, the UAV can implement the simulation of devices and their surrounding environments, carry out the data exchange between devices and UAV, execute the adaptation evolvement motivated by non-stationary environments, bring out smart trajectory design for itself and so on. 

% Due to the relationship between IoT networks and wireless cellular networks, the design of a UAV based DT system also attracted some attention in past years. 
% To be specific, the mobility design of the UAV can be difficult when both the intelligent service UAV trajectory is considered. 

\vspace{-0.1cm}

\section{System Model and Problem Formulation}\label{sec:Sys mdl}
% In this section, we present the considered system, including IoT devices and UAVs. 

\subsection{AoI Model for IoT Devices}
\begin{figure}[htbp]
	\centering
	\includegraphics[width=3.5 in]{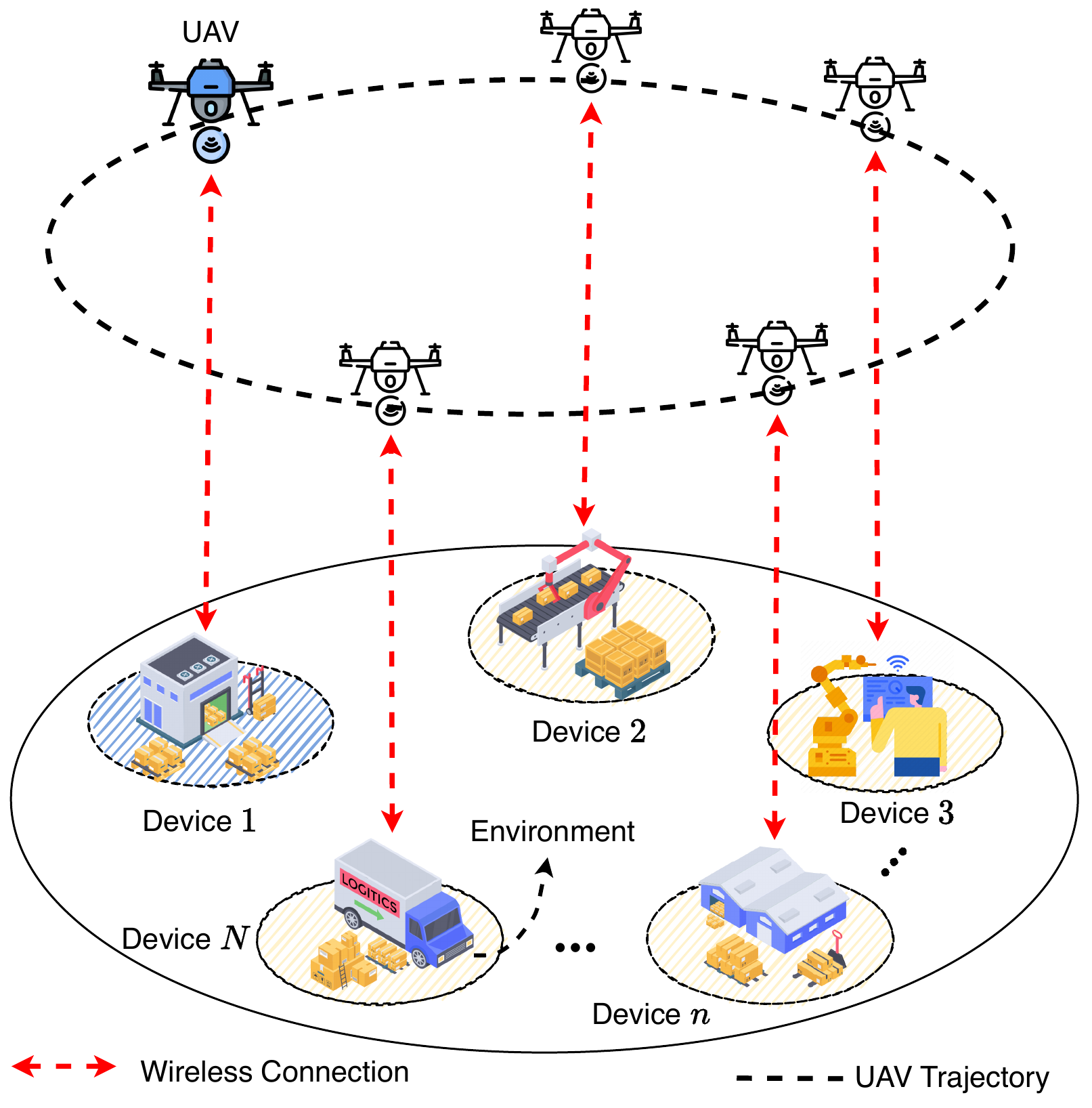}
	\caption{\small An illustration of the system model comprising of a UAV that sequentially visits the IoT devices which are distributed in a real-world non-stationary environment.}
	\label{System_UAV}
\end{figure}

% Consider an area with $N$ IoT devices, indexed by the set $\mathcal{N}$. The devices are randomly and uniformly distributed in the area, and each device has limited battery energy. The coordinates of device $i$ are $(x_i, y_i)$, $i \in \mathcal{N}$. Equipped with radio transceivers, microcontrollers, and sensors, the IoT devices can monitor and interact with their local environments~\cite{stoyanova2020survey}. 
% We assume that the local environments of the IoT devices are independent. 
% Consider a slotted system with index $t$ representing the time slots. The system operates over an infinite time horizon, i.e., $t = 0, 1, 2, \cdots$.
% The system may have many potential applications, such as precision agriculture, automatic traffic control, and smart monitoring. 
Consider a geographical area in which a set $\mathcal{N}$ of $N$ IoT devices are deployed as in Fig.~\ref{System_UAV}.
% There are $N$ IoT devices in the considered area. 
% The indexes of the devices are collected by the set $\mathcal{N}$. 
These devices are randomly distributed in this area according to a uniform distribution. 
The IoT devices are equipped with radio transceivers, microcontrollers, and sensors, enabling them to monitor and interact with their physical systems and surrounding environments~\cite{stoyanova2020survey}. 
For example, IoT devices in a smart factory can be deployed to monitor the production line, such as the equipments, products, etc. and their environmental conditions. 
Upon capturing the data, the IoT devices employ processing and data analysis to extract valuable insights, such as the equipment malfunctions, product quality, or any other useful events. 
In response to these detected events, each IoT device is required to take actions (e.g. trigger an alert, adjust the manufacturing parameters etc.).
% The frequency and complexity of these events can vary over time depending on several factors, including the fluctuation in raw material quality, production volume, machine health, shifts in customer demand and so on. 
These actions are time-sensitive and require real-time responses, which necessitates striking a balance between time criticality and energy consumption of IoT devices to maintain operational efficiency and sustainability.
% These actions are time-sensitive and often require immediate action response. This brings a balance between the urgency of event response and energy consumption of IoT devices which must be carefully designed to maintain the operational efficiency and sustainability. 
% Another smart farming example can be found in conference paper \cite{gong2021lifelong}.

The coordinates of each device $i$ are denoted as $\boldsymbol{d}_i = (x_i, y_i)$, where $i = 1, \ldots, N$. 
Furthermore, the system operates over a time horizon that is divided into equal time slots $t$, where $t = 1, 2, \ldots, T$, where $T$ is the total number of time slots.
At the beginning of each time slot $t$, device $i$ collects data from its environment. Here, the environment refers to the physical system and its surroundings.
We assume that the local environments of the IoT devices are independent from each other. 
The data packets that arrive at each device are independent and identically distributed (i.i.d.). Assume that the arrivals of data packets at device $i$ follow a Bernoulli distribution with a probability $\lambda_{i,t}$ at time slot $t$. 
Let $q_{i,t} \in \{0,1\}$ represent the arrival of a data packet of device $i$ at time slot $t$, where $q_{i,t} = 1$ indicates the arrival of a data packet to device $i$; and $q_{i,t} = 0$, otherwise.
% $\lambda_{i,t} = 1$, if a new data packet arrives at device $i$ at the beginning of time slot $t$; $\lambda_{i,t} = 0$, otherwise.
% $(q_{i,t} = 1$ if a new data packet arrives at device, $q = 1)$  otherwise.  
Also, assume that the size $a_{i,t} \ge 0$ of a data packet follows a Gaussian distribution with parameters $(\bar{a}_{i,t}, \sigma_{i,t}^2)$, where $\bar{a}_{i,t}$ is the average number of CPU cycles required to process a packet, and $\sigma_{i,t}$ is the standard deviation \cite{gong2021lifelong}.

At each time slot $t$, each device $i$ allocates a certain number of CPU cycles $\epsilon_{i,t} \in [0, \epsilon_{i,\textrm{max}}]$ for processing the received packets, where $\epsilon_{i,\textrm{max}}$ is the maximum number of CPU cycles of device $i$ per time slot. 
Given the constant length of the timeslots, the energy consumption per timeslot is simplified to become directly proportional to $\kappa_i \epsilon_{i,t}^3$ \cite{chandrakasan1992low}, where $\kappa_i$ is a paramater related to the chip architecture of the CPU.
% Since the energy consumed by a CPU cycle in each time slot grows quadratically with the CPU frequency \cite{chandrakasan1992low} for one second, which is the length of the time slot, the energy consumption of device $i$ is $\kappa_i \epsilon_{i,t}^3$, where $\kappa_i$ is a chip architecture related parameter.
%since the energy consumption per CPU cycle grows quadratically with the CPU frequency~\cite{}.
A first-come-first-serve (FCFS) policy is employed. At the end of slot $t$, 
% the number of CPU cycles that device $i$ needs to process the remaining packets in its buffer will be given by: 
the number of CPU cycles required to process the remaining packets in the queue of device $i$ is given by:
\begin{equation}
    b_{i,t+1} = \max\{b_{i,t} + q_{i,t} a_{i, t} - \epsilon_{i,t}, 0\} ,
\end{equation}
where $b_{i,t}$ is the total amount of data in the queue.

% We use AoI to measure the freshness of the information contained in the data packets. 
% The AoI is the time elapsed since the generation of the last received status update packet~\cite{kaul2012real}.
% $\Delta_{i, t}$ denotes the AoI of device $i$ at time slot $t$.
% Let $\xi_{i,t} \in \mathbb{N}$ be the index to the most recently processed data packet at slot $t$. 
% The generation time of packet $\xi_{i,t}$ is denoted by $u_{i, \xi_{i,t}}$.
% At time slot $t$, the AoI of the packet is $\Delta_{i, t} = t - u_{i, \xi_{i,t}}$.
% For each IoT device, the evolution of the AoI is given by
% \begin{equation}
% \Delta_{i,t+1} =
% \begin{cases}
% \Delta_{i,t} + 1 \, , & \textrm{if} \quad \omega_{i,t+1} = 0 \, , \\
% (t+1)- u_{i,\xi_{i,t+1}} \, ,& \textrm{if} \quad \omega_{i,t+1} =1 \, ,\\
% \end{cases}
% \label{AoI update}
% \end{equation}
% where $\omega_{i,t+1} \in \{0,1\}$. $\omega_{i,t+1} =1$, if the CPU cycles required for one data packet were all completed in the previous time slot; or $\omega_{i,t+1} =0$, otherwise. 
% We use the age of information (AoI) to measure the freshness of the information contained in the data packets. The AoI is defined as the time elapsed since the generation of the last status update packet received~\cite{kaul2012real}. Let $\Delta_{i, t}$ denote the AoI of device $i$ at time slot $t$, and $\xi_{i,t} \in \mathbb{N}$ index the most recently processed data packet at slot $t$. The generation time of packet $\xi_{i,t}$ is denoted by $u_{i, \xi_{i,t}}$. At time slot $t$, the AoI of the packet is given by $\Delta_{i, t} = t - u_{i, \xi_{i,t}}$.
To measure the freshness of information in the data packets, we consider the AoI at each device $i$
% It is defined as the time that has elapsed since the generation of the last status update packet received by a device~\cite{kaul2012real}. 
is $\Delta_{i, t} = t - u_{\xi_{i,t}}$ during time slot $t$, where $\xi_{i,t}$ is the index of the most recently processed data packet at device $i$ in time slot $t$ and $u_{\xi_{i,t}}$ is the time at which the most recent data packet at device $i$ was generated ~\cite{kaul2012real}.
Then, the evolution of the AoI for each device $i$ is stated as: 
\begin{equation}
\Delta_{i,t+1} =
\begin{cases}
\Delta_{i,t} + 1 \, , & \textrm{if} \quad \omega_{i,t+1} = 0 \, , \\
(t+1)- u_{\xi_{i,t+1}} \, ,& \textrm{if} \quad \omega_{i,t+1} =1 \, ,\\
\end{cases}
\label{AoI update}
\end{equation}
% where $\omega_{i,t+1} \in \{0,1\}$. 
where $\omega_{i,t+1} = 1$ if the number of CPU cycles required to process the data packets at time $t$ is sufficient (i.e., the queue is empty)  to complete computing; $\omega_{i,t+1} =0$, otherwise.

% If the number of CPU cycles required for a data packet was all completed in the previous time slot, then $\omega_{i,t+1} =1$; otherwise, $\omega_{i,t+1} =0$.

% \subsection{Age of Information (AoI)}

\begin{figure}
	\centering
	\includegraphics[width=3.5 in]{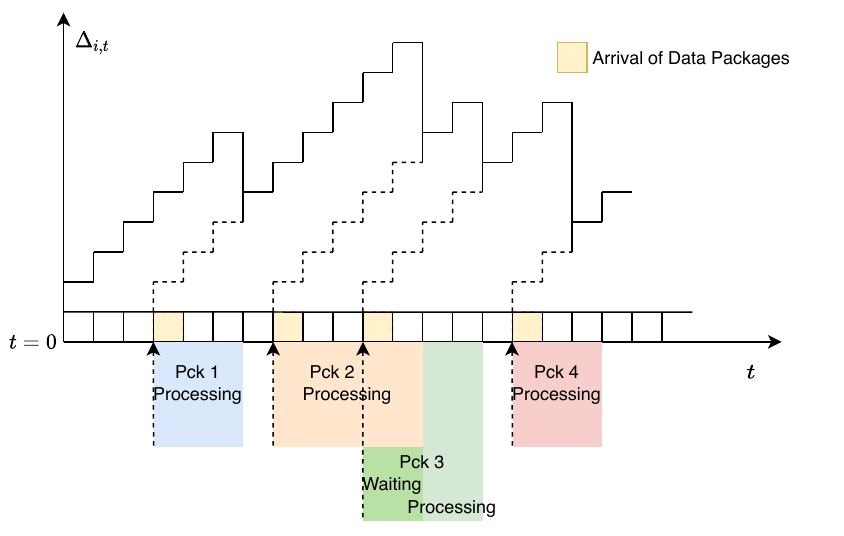}
	\caption{\small The dynamic evolution of AoI for device $i$. }
	\label{AoI_Sawtooth}
\end{figure}

To further elucidate this concept, Fig.~\ref{AoI_Sawtooth} illustrates the evolution of the AoI over time. Here, each device collects and packetizes data from its environment instantly. Upon its arrival at the IoT device, a data packet is processed immediately if there are no other packets in the queue. If there are packets ahead in the queue, the newly arrived packet incurs a queuing delay. This causes the AoI to increase until the newly arrived packet is processed, resulting in the sawtooth shape in Fig.~\ref{AoI_Sawtooth}.
Clearly, there exists a tradeoff between the AoI and energy consumption.
In other words, the more CPU cycles are allocated, the more data packets can be processed. 
Thus, this leads to better AoI performance, i.e., lower AoI.
% The higher the energy consumption is, the shorter AoI can be achieved. 
%Conversely, in order to reduce energy consumption, it leads inevitably to poor information freshness.
% As previously mentioned, there is a trade-off between the freshness of information and the energy consumption of the processing unit that must be captured in the cost function of the IoT devices. 
% Thus, the 
Henceforth, we define a cost function for each device $i$ to capture this tradeoff, as follows:
% A cost function is defined to capture the trade-off.
% At time slot $t$, the cost of device $i$, denoted by $c_i(t)$, is written as
\begin{equation}
c_i(t) = \beta \Delta_{i,t} + (1 - \beta) \kappa_i \epsilon_{i,t}^3 ,
\label{original_cost}
\end{equation}
where $\beta \in [0,1]$ is a 
% pre-configurable coefficient. 
factor to balance the tradeoff between the AoI and energy consumption during each time slot $t$.

\vspace*{-0.32cm}

\subsection{Non-Stationary Environment Model}

As a result to the dynamic changes over time, 
% (e.g., changes in weather, temperature, etc.) 
the environment experienced by each IoT device varies in a non-stationary fashion. These variations are reflected in the probability of data packet arrivals and the distribution of the sizes of the arrived data packets.
Thus, each IoT device faces challenges in adapting its CPU cycles allocation accordingly to its new environment.
This entails dynamically adjusting the allocation strategies to meet the varying requirements and optimize overall system performance.

\begin{definition}
	An \emph{environment} corresponds to a unique set of environmental parameters.
	The environment of device $i$ at time slot $t$ is represented by 
	$\boldsymbol{z}_{i,t} = ( \lambda_{i,t}, \bar{a}_{i,t}, \sigma_{i,t}^2, \kappa_{i}, \epsilon_{i,\textrm{max}})$. 
	% $j \in \mathbb{N}$.
	%  where $\mathbb{N}$ is the set of natural numbers.
 \label{Definition_Environment}
\end{definition}

In the environment tuple, $\kappa_{i}$ and $\epsilon_{i,\textrm{max}}$ remain constant for each device but may vary across different devices. These parameters represent the physical properties of each device that are inherent to the device itself and cannot be altered by the surrounding environments.
However, since $\kappa_{i}$ and $\epsilon_{i,\textrm{max}}$ impact the CPU allocation strategies, it is reasonable to include them as part of the environment-specific tuple.

Furthermore, we assume that each IoT device $i$ experiences $n_i$ environments within the time interval $[0, T)$. Thus, $Z = \sum_{i=1}^{N} n_i$ is the total number of environments experienced by all the devices. However, it is worth noting that the value of $n_i$ and, consequently, $Z$ are unknown in advance.
We assume that the intra-environment variations are stable over time while the inter-environment is non-stationary. 
% To simplify the analysis further, we assume that t
The abrupt changes of the environments occur at the beginning of each time slot. 
% In this scenario, each environment has a specific duration or period. 
The average duration of an environment for device $i$ is $p_i$, which follows a Gaussian distribution with average $\bar{p}_i$ and variation $\sigma'_i$. Indeed, determining the exact distribution of $p_i$ in advance can be challenging.
% $ P_i \sim \mathcal{N} (\text{Avg}(P_i), \text{Var}(P_i))$.
In addition, the average duration of the environments remains consistent for each device, since
% . This assumption comes from the fact that different devices may experience varying surrounding environments, but 
the environments encountered by a single device follow a certain pattern or regulation.

\vspace{-0.32cm}

\subsection{UAV Model}
% %A UAV serves as a flying base station to help devices in non-stationary environments. Without loss of generality, the UAV starts from a known starting point $\boldsymbol{l}_0= (0, 0)$. The UAV's location is $\boldsymbol{l}_m = (x_i, y_i)$, where $m$ means this is the $m$-th flying decision. It is assumed that the UAV would fly $M$ times during the considered time horizon. $M$ varies with the dynamic environments of IoT devices. A constant velocity $v_{\textrm{min}} \le v_m \le v_{\textrm{max}}$ is assumed when the UAV flies from one device to another. $v_{\textrm{min}}$ and $v_{\textrm{max}}$ are the UAV's minimum and maximum velocities, respectively. 
% A UAV serves as a flying base station to provide connectivity to devices in non-stationary environments. The UAV starts at a known location $\boldsymbol{l}_0= (0, 0)$ and flies to various locations $\boldsymbol{l}_m = (x_m, y_m)$ over a time horizon, {\color{red}where $m$ is the number of flights made by the UAV. The number of flights, $M$, can vary depending on the changing need for IoT devices. The UAV is assumed to maintain a constant velocity $v_{\textrm{min}} \le v_m \le v_{\textrm{max}}$, as it moves from one device to another, where $v_{\textrm{min}}$ and $v_{\textrm{max}}$ are the minimum and maximum velocities of the UAV, respectively.}
A UAV serves as a flying BS for IoT devices.
The proposed approach involves the UAV flying from one device to another to collect environment-related data. 
This data collection process enables the UAV to assist the devices in adapting to the dynamic environments it encounters.
% The UAV flies from one device to another to collect environment related data and then help the device adapting to dynamical environments.
% connect the devices in the non-stationary environments. 

Initially, the UAV starts at a known location $\boldsymbol{l}_0$. We consider $m = 1, \ldots, M$ to indicate the $m$-th flight made by the UAV, where $M$ is the total number of flights that the UAV can make during $T$ time slots. Then, we define $\boldsymbol{l}_m \in \{\boldsymbol{d}_i\} $, $\forall m \in \mathcal{M}\triangleq  \{1,\ldots, M\}$.
However, the exact value of $M$ is considered to be unknown.
The UAV is assumed to maintain a constant velocity $v_{\textrm{min}} \le v_m \le v_{\textrm{max}}$ when flying from one device to another, where $v_{\textrm{min}}$ and $v_{\textrm{max}}$ are the minimum and maximum velocities of the UAV, respectively.
In addition, once the UAV arrives at a device, it will hover over it before it flies to the next destination. The distance between UAV and the device is considered close enough such that the energy consumed for their communication can be neglected.
Moreover, we assume that the velocity of the UAV permits it to visit each environment for multiple times.

%When flying from $\boldsymbol{l}_m$ to $\boldsymbol{l}_{m+1}$, the total energy consumption of the UAV, denoted by $e^u_m$, consists of propulsion energy, communication energy \cite{fotouhi2019survey, zeng2017energy}, and computing energy. The communication energy and computing energy are relatively low, compared to the propulsion energy \cite{zeng2019energy}. 
When flying from one location $\boldsymbol{l}_m$ to another location $\boldsymbol{l}_{m+1}$, the UAV consumes energy for propulsion, communication, and computing. 
Here, assume a rotary wing UAV whose total energy $e^U_m$ is dominated by propulsion~\cite{ zeng2019energy}.
% Consider a rotary-wing UAV. 
The energy of the UAV can be expressed as~\cite{zeng2019energy}:
\begin{equation}
\small
e_0 (v_m)\! = \! 
\underbrace{     P_0 \left(  \! 1 + \dfrac{3v_m^2}{v_{\textrm{tip}}^2}   \!\right)  }_{\textrm{blade\ profile}}   
\!+ \!\underbrace{        P_i \left( \!  \sqrt{1+\dfrac{v_m^4}{4 v_0^2}}   \! - \!  \dfrac{v_m^2}{2v_0^2}     \right) ^{1/2}        }_{\textrm{induced}}  \\
\!+   \!   \underbrace{   \dfrac{1}{2} d_0 \rho s A v_m^3    }_{\textrm{parasite}} \, ,
\label{UAVFlying}
% \normalsize
\end{equation}
where $P_0$ and $P_i$ denote the blade profile power and induced power, respectively, $v_{\textrm{tip}}$ is the tip speed of the rotor blade, $v_0$ is the mean rotor-induced velocity in hover, $d_0$ is the fuselage drag ratio, $s$ is the rotor solidity, $\rho$ is the air density, and $A$ is the rotor disc area. 
Hence, the total energy consumed by the UAV during its $m$-th visit can be formulated as:
\begin{equation}
e^U_m  (v_m, \boldsymbol{l}_m, \boldsymbol{l}_{m+1} )= \dfrac{ \|\boldsymbol{l}_{m+1} - \boldsymbol{l}_m\|  }{v_m} e_0 (v_m) \, .
\end{equation}
% where $\|\boldsymbol{l}_{m+1} - \boldsymbol{l}_m\|$ is the Euclidean distance between the locations $\boldsymbol{l}_m$ and $\boldsymbol{l}_{m+1}$. 

% Similarly, when the UAV is hovering, its energy consumption can be given by
% \begin{equation}
% e^h_m(q_{m, \textrm{in}}, q_{m, \textrm{out}}) = (q_{m, \textrm{out}} - q_{m, \textrm{in}}) \cdot e_0 (v_{\textrm{min}}) .
% \end{equation}
%\textcolor{red}{During a hovering time $(q_{m, \textrm{out}} - q_{m, \textrm{in}})$, we have a constant energy $e^h_m(q_{m, \textrm{in}}, q_{m, \textrm{out}})$.}

% The total energy consumption for UAV's $m$ th visit can be obtained as:
% \begin{equation}
% e^u_m = e^f_m (v_m, \boldsymbol{l}_m, \boldsymbol{l}_{m+1} ) + e^h_m(q_{m, \textrm{in}}, q_{m, \textrm{out}}) \, .
% \end{equation}

%Energy Related: \\
%$e^f_m (v_m, \boldsymbol{l}_m, \boldsymbol{l}_{m+1} )$: flying energy from $\boldsymbol{l}_m$ to $\boldsymbol{l}_{m+1} $ \\
%$e^h_m(u_m, q_m)$: hovering energy over location $\boldsymbol{l}_m$ \\
%$e^c_m$: constant margin energy, such as wireless transmission and processing \\ %consumption, which is assumed to be a constant\\
%$e^u_m$: total energy consumption of UAV's $m_{th}$ visit \\

Here, we consider two types of transmissions: a) uplink data collection and b) downlink strategy update. However, as a small amount of sampled data is typically uploaded, we can neglect the time and energy required for uplink data transmission \cite{zhou2011modeling}. 
Similarly, as the downlink transmission involves the transmission of strategies for IoT devices, it is reasonable to disregard this time and energy as well \cite{wu2016green}. This is due to the fact that the strategies of all the devices have the same structure.
% With respect to the processing time and energy consumption of the UAV, we emphasize the strategy update procedure and UAV's hovering consumption, respectively. 
In addition, we consider that the processing time and energy consumption of the UAV are mainly governed by the strategy update procedure and hovering, respectively.
Based on the upcoming display of the strategy update process, it is clear that each strategy update involves similar steps and consumes the same amount of energy. Hence, it is valid to consider it as a constant within in each flight. 
Leveraging the fact that the strategy update procedure is similar and the distance between the UAV and the device is near, we assume the hovering time is the same.
Since the hovering energy consumed by the UAV is proportional to the number of time slots it hovers \cite{zeng2019energy}, the hovering time and hovering energy can be treated as constant values.
% Similarly, the processing steps involved in each flight are similar and consistent. As such, it is reasonable to treat the hovering time and hovering energy as constant values as well.
% As such, the energy consumed can be taken as a constant value. 
As this constant on each flight would not impact the optimization process, it can be ignored in the optimization.
Therefore, the total energy consumed by the UAV during each flight is equivalent to $e^U_m$.

\vspace*{-0.32cm}

\subsection{Problem Formulation}\label{sec:Prob State}
Our goal is to minimize the cost for all devices and the energy consumption of the UAV. The problem can be formulated as:
\begin{alignat}{2}
\min_{ \boldsymbol{E}, \boldsymbol{v}, \boldsymbol{F} }  &    \dfrac{1}{Z} \!   \sum_{i=1}^{N}  \sum_{t=0}^{T}  \mathbb{E}_{\lambda_{i,t}, a_{i,t}} [ c_{i}(t) ] \!  +\!  \dfrac{\mu}{M} \!  \sum_{m=0}^{M} \! e^U_m  (v_m, \, \boldsymbol{l}_m, \, \boldsymbol{l}_{m+1} ) \label{Problem_1} \\
\mathrm{s.t.} \quad 
&\Delta_{i,t} \in \mathbb{N}, \quad  \forall i \in \mathbb{N},  \; t \in [0, T) , \label{eqn1} \\
& \epsilon_{i,t} \le \epsilon_{i,\textrm{max}},  \forall i \in \mathbb{N}, \; t \in [0, T) ,       \label{eqn2} \\
& v_{\textrm{min}} \le v_m \le v_{\textrm{max}}, \: \forall \, m= 0, 1, \ldots, M ,  \label{eqn3} \\ 
% & \sum_{j = 1}^{n_i} (t_{i,j+1} - t_{i,j}) = T, \quad \forall i \in \mathbb{N},  \label{eqn4}  \\
& \boldsymbol{l}_m \in \{\boldsymbol{d}_i\}, \: \:   \forall m = 1, \dots, M, \, \forall i= 1, \ldots, N,    \quad  \label{eqn5} 
\end{alignat}
% where $Z = \sum_{i=1}^{N} n_i$ is the number of environments during $T$ time slots. 
% Here, $n_i$ is the number of different environments that device $i$ experiences during time $[0, T)$. 
where $\mu \in [0, 1]$ is a parameter that regulates the tradeoff between the devices' cost and the UAV energy usage.
In addition, $\boldsymbol{\varepsilon}_i = [\epsilon_{i,0}, \ldots, \epsilon_{i,T}]$ indicates the vector of CPU cycles of device $i$ throughout the period $T$,
$\boldsymbol{E} = [\boldsymbol{\varepsilon}; \ldots; \boldsymbol{\varepsilon}_N]$ is the matrix of CPU cycles for all devices,
$\boldsymbol{v} = [v_1, \ldots, v_M]$ is the vector of UAV's velocity during its flight, 
$ \boldsymbol{F} = [\boldsymbol{l}_1, \dots, \boldsymbol{l}_M]$ is vector of the UAV's destination during its flight.
%a $N \times T$
% Constraints \eqref{eqn2} and \eqref{eqn3} ensure that the UAV flies at a feasible velocity and that the total time of all environments experienced by a device is $T$, respectively. 
Constraints \eqref{eqn2} and \eqref{eqn3} indicate the limits of the CPU cycles of each device and the UAV, respectively.
Constraint \eqref{eqn5} specifies that the target point of a UAV flight is the location of the selected device. The UAV flies in a straight line to the target point.
It is worth noting here that the distribution of the data packet arrival $\lambda_{i,t}$ and the distribution of the sizes of the arrived packet $a_{i,t}$ for each device $i$ in time slot $t$ are unknown.
In addition, the duration of any environment follows an unknown distribution.
% , making it unpredictable as in real-life scenarios..

% $\mathcal{P}_1$ and $\mathcal{P}_2$ represent the unknown distribution of the data package arrival probability and distribution of the size of the arrived packages for each device throughout the time period, retrospectively. It is worth noting that the lasting time of any environment follows an unknown distribution $\mathcal{P}_{3}$ which makes them abrunpt and unpredictable, where $t^s_{i,t}$ is the starting time of any environment for device $i$ and $t^e_{i,t}$ is the ending time.
% We assume the length of the environments are uncertain and unpredictable such that $(t^e_i - t^s_i) \sim \mathcal{P}_{3}$, where $t^s_i$ is the starting time of any environment for device $i$, $t^e_i$ is the ending time and $\mathcal{P}_{3}$ is the unknown distribution of length of each environment for all devices. This indicates that the duaration of each environment is uncertain and unpredictable.

\vspace*{-0.42cm}

Furthermore, the UAV assists the devices in a sequential manner. 
When the UAV arrives at an IoT device during its $m$-th flight, it collects the environment related information from the device and updates its strategy. This strategy reflects the device's interaction with its current environment. In particular, it helps the device determine CPU cycle allocation at each time slot before the UAV comes for its next visit. 
% The interaction strategy is denoted by $\mathcal{E} = \{ \epsilon_{i,j,t} \}$. 
This can help the device improve its decisions in its current environment. 
% before a new environment appears.

To help the devices update their interacting strategies, the UAV also learns its own trajectory. 
It determines its flight path $\boldsymbol{F}$ and $\boldsymbol{v}$.
%  for all $m \in \mathcal{M}$, $i \in \mathcal{N}$.
On the one hand, the decisions made by the UAV on the flight path $\boldsymbol{F}$ and velocity $\boldsymbol{v}$ can have a strong impact on the first term in the objective \eqref{Problem_1}. 
% Different flight paths and velocities can affect the time it takes for the UAV to reach a device, and arriving 
In essence, early or late visits of the UAV at a device can affect the cost of the device, particularly in the face of unknown changes in the environment.
For instance, we can consider the scenario of a delayed visit. If the UAV visits a device significantly late after the device has already encountered a new environment, the device may acquaint an outdated and unsuitable interacting strategy in the absence of the UAV. Due to changes in data packet arrival probability and size, this inappropriate strategy can result in elevated AoI and unnecessary CPU energy consumption. Consequently, the arrival time, i.e., the flight path of the UAV, can have a vital impact on the first term in \eqref{Problem_1}.
On the other hand, the decision making process of the devices, i.e., $\boldsymbol{E}$,
also affects the UAV's energy consumption, i.e., the second term in \eqref{Problem_1}.
% as indicated by the objective function in~\eqref{Problem_1}. 
 For instance, if a device's CPU decisions are suboptimal, i.e., its interaction strategy is poor, this can lead to repetitive UAV visits for collecting data or extended the hovering times over the device for data collection. Consequently, the probability that the UAV visits other devices with new environments decreases. This, in turn, can degrade the overall performance of the entire system.
Hence, in either case, 
% Both of these scenarios can increase 
the energy consumption of the UAV increases.

Furthermore, the flight destination of the UAV $\boldsymbol{l}_m$ in \eqref{Problem_1} is discrete. Hence, the problem is a Mixed Integer Programming (MIP) problem. As this problem is NP-hard, finding an optimal solution for the problem can be highly challenging.
%To solve the optimization problem, we decompose it into two parts a: the UAV's decisions and the devices' decisions, and iteratively solve the two parts in an alternating manner. Specifically, we fix the UAV's flying decisions as $\mathcal{L}_\Theta$ and $\mathcal{V}_\Theta$ and optimize $\mathcal{E}$. Then, we determine the optimal UAV flying decisions $\mathcal{L}^* $ and $\mathcal{V}^*$ given the optimized $\mathcal{E}^* = {\epsilon^*_{i,j,t}}$.
To efficiently solve this problem, we decouple it into two sub-problems that are solved in an alternating manner. First, we fix the flight control of the UAV
% denoted by $\mathcal{L}_\Theta$ and $\mathcal{V}_\Theta$, 
and optimize the decision-making processes of the IoT devices.
% $\mathcal{E}$. 
Up to this end, we resort to a lifelong learning solution as the environments of these devices are considerably non-stationary. 
The obtained optimal decisions are denoted by $\boldsymbol{E}^*$, consisting of optimal CPU cycles decision $\epsilon_{i,t}^*$ for all devices and all time slots.
%Then, using the optimized $\mathcal{E}^* = \{\epsilon^*_{i,j,t}\}$, 
Subsequently, we utilize the optimal interaction strategies obtained in the first stage to calculate the UAV's optimal flying decisions using an AC network.
% Then, we will use the obtained optimal interaction strategies in the first stage to compute the optimal flying decisions of the UAV, 
% We determine the optimal flight control decisions for the UAV, 
Thus, the obtained optimal flying decisions are denoted as $\boldsymbol{F}^* = [\boldsymbol{l}_1^*, \dots, \boldsymbol{l}_M^*]$ and $\boldsymbol{v}^* = [v_1^*, \ldots, v_M^*]$.

\section{Lifelong RL for IoT Strategy Optimization in Non-stationary Environments}\label{sec:LL slution}

The first part of the solution emphasizes on the first term in \eqref{Problem_1}. 
As such, the flight path and flying velocities of the UAV are considered to be constant throughout the time period $T$. Hence, the optimization problem~\eqref{Problem_1} is reduced to the following AoI-energy cost minimization problem via optimizing $\boldsymbol{E}$, that can be formulated as follows:
% As we declared above, the flight path and flying velocities of the UAV will be fixed while we optimize the first term. 
% As such, problem~\eqref{Problem_1} can be rewritten as:
% Given fixed flight control $\mathcal{L}_\Theta$ and $\mathcal{V}_\Theta$, Problem~\eqref{Problem_1} is rewritten as
%写成公式(7)
\begin{alignat}{2}
	\min_{ \boldsymbol{E}   }  &   \quad    \dfrac{1}{Z}    \sum_{i=1}^{N}  \sum_{t=0}^{T}  \mathbb{E}_{\lambda_{i,t}, a_{i,t}} [ c_{i}(t) ]   
% +  C_1  
\label{Problem_1_LV} \\
	\mathrm{s.t.} \quad 
	& \eqref{eqn1}, \eqref{eqn2}. 	\nonumber 
\end{alignat} 
% in addition, $v_m = v_\Theta, \forall  m= 0, 1, \cdots, M $, is a constant value that is adopted by the UAV throughout the time period $T$. 
During each flight, the UAV destination is randomly selected from the pool of all IoT devices.
%  such that $\boldsymbol{l}_m = \boldsymbol{d}_i, \: i =  \textrm{random}(1, \ldots, N)$.
% Here, the UAV's velocity is fixed at $v_\Theta$ for the duration of the flight. The UAV follows a predetermined flying pattern, $\mathcal{L}_\Theta $, in which it starts at the starting point $\boldsymbol{l}_0 = (0, 0)$ and visits the devices in the order specified by $\mathcal{N}$. 
% $C_1$ in \eqref{eq: constant C1} is a constant calculated from $\mathcal{L}_\Theta$ and $\mathcal{V}_\Theta$, and can be suppressed in the problem. 
The problem in \eqref{Problem_1_LV} involves optimizing the performance of IoT devices in an environment that undergoes discrete changes in an unpredictable manner. 
% These changes occur in a discrete and independent manner over a time period denoted as $T$. 
% As previously mentioned, these non-stationary environments are defined by a tuple $\emph{z}_{i,t}$.
In other words, each environment
% can be represented by a unique tuple $\emph{z}_{i,t}$ that 
persists for an unknown duration before being replaced by a new one. 
Furthermore, each environment can be considered as a stationary environment characterized by the tuple $\emph{z}_{i,t}$ throughout its duration. Consequently, the entire problem described in \eqref{Problem_1_LV} can be viewed as a collection of independent environments that appear sequentially.

% To solve this problem, we inspect several prospective methods.
Here, traditional optimization approaches such as convex and stochastic optimization are unsuitable to solve this problem as they rely on stationary environments in which the distribution of the generated data is known. 
However, in our case, the dynamical patterns of the environments are unknown and non-stationary.
Moreover, classical machine learning algorithms such as supervised learning and RL are also unsuitable for solving this problem as the setting does not take place in a stationary condition. 
% As a method that can learn knowledge from one source domain and transfer it to the target domain, transfer learning (TL) does not fit in our situation. 
% This is because domain similarities, which are required between the source and target domains, cannot be assured in our unpredictable environments. Vastly different domains can harm the performance with negative transfer.
In addition, a continuous knowledge transfer method is needed to leverage the acquired knowledge from one environment to the other. 
% multi-task learning (MTL) aims to improve the performance of multiple related tasks by jointly learning them. However, instead of learning in a sequential way, the MTL learns shared information across tasks simultaneously. This is not applicable to our case where all the environments are not available at the beginning of the learning process.
\begin{figure}%[htbp]
	\centerline
	{\includegraphics[width=3.3 in]{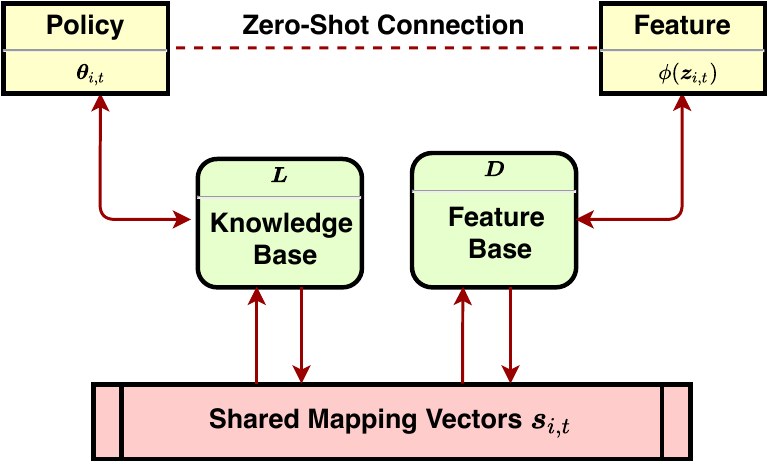}}
	\caption{\small An illustration of the relationship between the knowledge base and environment related vectors.}
	\label{LSD}
\end{figure}
Hence, a continual learning method that allows a sequential stream of tasks to be acquired should be considered.
Notably, lifelong learning has the potential to address the challenges presented by this problem. 
% Meta-learning involves learning how to learn for multiple tasks and can handle the uncertainties of learning by developing intuitions for new tasks. 
% As in \cite{finn2017model}, a continual learning method is proposed to solve the non-stationary environments.
%The concept of lifelong learning was originally introduced in \cite{thrun1995lifelong, silver2013lifelong}.
Despite the early proposal of lifelong learning in literature~\cite{thrun1995lifelong, silver2013lifelong}, remarkably, most works have been limited to supervised learning methods and were not exploited in the RL regime.
%Recently, the surge of RL has also propelled the advancement of lifelong learning. 
% This can greatly broaden the application domains of the lifelong learning. 
Hence, we propose an efficient lifelong RL approach where each environment can be modeled as a RL problem and knowledge can be transferred between them accordingly.

\subsection{RL for Independent Environments}{\label{IIIA}}

Now, we will first show the optimization model of each independent environment. Then, we will provide its corresponding MDP. Each independent environment can be modeled through the following problem:
\begin{alignat}{2}
	\min_{ \boldsymbol{E}   }  &   \quad        \mathbb{E}_{\lambda_{i,t}, a_{i,t}} [ c_{i}(t) ]   
% +  C_1  
\label{Problem_2_Single} \\
	\mathrm{s.t.} \quad 
	& \eqref{eqn1}, \eqref{eqn2}. \nonumber
	% \eqref{eqV}, \eqref{eqL}, \eqref{eqInterval},
	% \nonumber \\
	% & \lambda_{i,t} \in \{0,1\} , \\
	% & a_{i,t} \sim \mathcal{N}(\bar{a}_{i,t}, \sigma_{i,t}) , \forall i = 1, \ldots, N  \label{eqSingleDis} \, , \\
	% \,  t \in [t^s_{i,t}, t^e_{i,t}),  \nonumber
\end{alignat}
It is worth noting that the start time and end times of each environment in time slot $t$ are $t^s_{i,t}$ and $t^e_{i,t}$, respectively. In addition, we note that $0 \le t^s_{i,t} < t^e_{i,t} \le T$ while having these variables unknown in advance.
% where $0 \le t^s_{i,t} < t^e_{i,t} \le T$ and $t \in [t^s_{i,t}, t^e_{i,t})$ is the lasting time of the studied environment for device $i$.
% $t_s$ and $t_e$ are separately the starting time slot and ending time slot of an independent environment.
% that can be characterized as $\emph{z}_{i,t} =  ( \lambda_{i,t}, \bar{a}_{i,t}, \sigma_{i,t}^2, \kappa_{i}, \epsilon_{i,\textrm{max}})$.
% It is worth noting that $t^s_{i,t}$ and $t^s_{i,t}$ are unknown in advance.
% Other conditions like $v_m = v_\Theta$ and $\boldsymbol{l}_m = \boldsymbol{d}_i$ are satisfied as in \eqref{Problem_1_LV}.

Let $\chi_i$ refer to whether device $i$ experiences an environmental change, by having $\chi_{i} = 1$ indicating a new environment and $\chi_{i} = 0$ indicating that there is no change in the environment. 
For each environment, we can model the decision-making process of each device as a MDP, as delineated in the following.

In this system, each IoT device interacts with its surrounding environment at each time slot by processing data packets received from the environment. During this interaction, the device generates a record of its status changes, i.e., \emph{interaction history}, denoted by $\boldsymbol{\tau}$. The space in which this interaction history occurs is defined as $\{\boldsymbol{X}, \boldsymbol{Y}, \boldsymbol{R}\}$, where $\mathcal{X}$ is the state space, $\mathcal{Y}$ is the action space, and $\mathcal{R}$ is the reward function, and are elucidated as follows:

\begin{itemize}
    \item 

%The state space $\mathcal{X}  \subset \mathbb{R}^d$ is the set of AoI values and number of pending data cycles at the beginning of each time slot, i.e., $\mathcal{X} = \{\boldsymbol{x}_{i,j,t}\}=  \{  \Delta_{i,j,t}, b_{i,j,t}  | \Delta_{i,j,t} \in \mathbb{N}, b_{i,j,t} \in  \mathbb{N}\}$. Here, $d$ equals to the number of variables in state space. The action space is the set of all possible CPU cycles,  $\mathcal{Y} = \{\boldsymbol{y}_{i,j,t}\} = \{\epsilon_{i,j,t}| 0 \le \epsilon_{i,j,t} \le \epsilon_{i,\textrm{max}},  \epsilon_{i,j,t}  \in \mathbb{R}\}$. The reward function can be defined as $R(\boldsymbol{x}_{i,j,t}, \boldsymbol{y}_{i,j,t}) = - (\beta \Delta_{i,j,t} + (1-\beta) \kappa_i \epsilon_{i,j,t}^3  )$, which is the negative of the cost function \eqref{original_cost}. At each time slot, the decision-making of the $\boldsymbol{y}_{i,j,t}$ based on $\boldsymbol{x}_{i,j,t}$ can be essential to achieve cost minimization for each device $i$ in environment $j$. As such, the parameterized decision-making policies can be defined as $\Pi_{i,j} = \{ \pi_{{\boldsymbol{\theta}}_{i,j}} | \boldsymbol{\theta}_{i,j} \in \mathbb{R}^d \}$, where $\pi_{{\boldsymbol{\theta}}_{i,j}}(\boldsymbol{y}_{i,j,t} | \boldsymbol{x}_{i,j,t}) = \text{Pr}\{ \boldsymbol{y}_{i,j,t} | \boldsymbol{x}_{i,j,t}, \boldsymbol{\theta}_{i,j}  \} $.
The state space $\mathcal{X}$ is a set of tuples $\boldsymbol{x}_{i,t} = (\Delta_{i,t}, b_{i,t})$ representing the AoI and the number of CPU cycles required pending data packets at the beginning of slot $t$ for device $i$.
%  in environment~$j$.  
%$\Delta_{i,j,t} \in \mathbb{N}$ is the AoI. $b_{i,j,t} \in \mathbb{N}$ is the number of CPU cycles required pending data packets. 
The state space has a dimension of $d = 2$, which is the number of variables in the state space.

\item
The action space is the set of all possible CPU cycles $\mathcal{Y} = 
\{\epsilon_{i,t},\,\forall i,t\}$.
% where $0 \le \epsilon_{i,t} \le \epsilon_{i,\textrm{max}}$. 

\item
The reward function $\mathcal{R}(\boldsymbol{x}_{i,t}, \epsilon_{i,t})$ is 
% the negative of the cost function, as 
given by 
$$\mathcal{R}(\boldsymbol{x}_{i,t}, \epsilon_{i,t}) = - (\beta \Delta_{i,t} + (1-\beta) \kappa_i \epsilon_{i,t}^3).$$ 
% where $0 \leq \beta \leq 1$ and $\kappa_i$ are two constants specifying the relative importance of the AoI and CPU cycles in the cost function, respectively.

\item 
%The decision-making policies $\Pi_{i,j}$ are parameterized functions $\pi_{{\boldsymbol{\theta}}{i,j}}$ that determine the probability of selecting a particular action $\boldsymbol{y}{i,j,t}$ given a state $\boldsymbol{x}{i,j,t}$ and a set of parameters $\boldsymbol{\theta}{i,j} \in \mathbb{R}^d$. The decision-making policies are defined as $\Pi_{i,j} = { \pi_{{\boldsymbol{\theta}}{i,j}} | \boldsymbol{\theta}{i,j} \in \mathbb{R}^d }$, where $\pi_{{\boldsymbol{\theta}}{i,j}}(\boldsymbol{y}{i,j,t} | \boldsymbol{x}{i,j,t}) = \text{Pr}{ \boldsymbol{y}{i,j,t} | \boldsymbol{x}{i,j,t}, \boldsymbol{\theta}{i,j} }$. The goal is to find the optimal decision-making policy $\pi_{{\boldsymbol{\theta}}_{i,j}^*}$ that minimizes the cost function for each device $i$ in environment $j$.
$\Pi_{i,t}$ is defined as the set of policy parameters:
% The decision-making policies $\Pi_{i,t}$, $\forall i,$ are parameterized functions $\pi_{{\boldsymbol{\theta}}_{i,t}}$ that determine the probability of selecting a particular action $\boldsymbol{y}_{i,t}$, given state $\boldsymbol{x}_{i,t}$. 
% The set of policy parameters are denoted as $\boldsymbol{\theta}_{i,t} \in \mathbb{R}^d$; i.e.,  
$$\Pi_{i,t} = \{ \pi_{{\boldsymbol{\theta}}_{i,t}} | \boldsymbol{\theta}_{i,t} \in \mathbb{R}^d \},$$ 
where $\boldsymbol{\theta}_{i,t}$ is the policy for device $i$ at time slot $t$. 
In addition, $\pi_{{\boldsymbol{\theta}}_{i,t}}(\epsilon_{i,t} | \boldsymbol{x}_{i,t}) = \Pr\{ \epsilon_{i,t} | \boldsymbol{x}_{i,t}, \boldsymbol{\theta}_{i,t} \}$ is the parameterized function that determines the probability of selecting a particular action $\epsilon_{i,t}$, given state $\boldsymbol{x}_{i,t}$.  The goal is to find the optimal decision-making policy $\pi_{{\boldsymbol{\theta}}_{i,t}^*}$ that minimizes the cost function for device $i$ in the current environment.
\end{itemize}

%Since each environment has a unique $\boldsymbol{z}_{i,t}$. 
Furthermore, each environmental MDP episode can be equivalently represented through a one-to-one mapping utilizing the variables of $\boldsymbol{z}_{i,t}$.
% such as $\lambda_{i,t}$, $a_{i,t}$, the device's attributes $\kappa_{i}$ and $\epsilon_{i,\textrm{max}}$. 
% as an equivalent expression for an environment.
Subsequently, we use an environment feature vector $\phi(\boldsymbol{z}_{i,t}) \in \mathbb{R}^{d_z}$ to represent each unique MDP episode.
%\footnote{Here, we note that "environmental MDP episode" and "MDP episode" are used interchangeably in the rest of the paper.}. 
$\phi(\cdot)$ is a feature extraction function, 
% that can be linear or non-linear, 
and $d_z$ is its dimension. While multiple features can describe the same environment, different environments typically have unique features. However, the exact environment descriptor $\boldsymbol{z}_{i,t}$ is unknown and must be estimated from the collected interaction history $\boldsymbol{\tau}$.

Up so far, we have acknowledged the independence between environments. 
We proceed to determine the environment change detection. 
% However, the issue of determining the boundary points between these environments remains unresolved. 
It is eligible to consider that the change of $\boldsymbol{z}_{i,t}$ indicates the change of environments.
As such, $\boldsymbol{z}_{i,t}$ can be detected by sampling and collecting the environment related information. The environment related information can be extracted from interaction history $\boldsymbol{\tau}$.
% Given a set of interaction history  $\boldsymbol{\tau} = \{\boldsymbol{x}_{i,j,t}, \boldsymbol{y}_{i,j,t}, R(\boldsymbol{x}_{i,j,t}, \boldsymbol{y}_{i,j,t})\}$, $T_{i,j}$ is the length of the collected interaction history, let $\mathcal{Q}_{i,j} = \{t \in \mathbb{N}| b_{i,j,t+1} +\epsilon_{i,j,t} >  b_{i,j,t}\}$ be the set of time slots at which there was a new packet arrival.
% Hereinafter, the task descriptor can be identified as: 
% $\tilde{\lambda}_{i,j} \approx Q_{i,j}/T_{i,j}$, 
% $\tilde{a}_{i,j} \approx 1/Q_{i,j} \sum_t (b_{i,j,t+1} - b_{i,j,t})$, 
% $\tilde{\sigma}^2_{i,j} \approx 1/Q_{i,j} \sum_t (b_{i,j,t+1} - b_{i,j,t} - \bar{a}_{i,j}) $,
% where $t \in \mathcal{Q}_{i,j}$.
Given interaction history $\boldsymbol{\tau} = \{\boldsymbol{x}_{i,t}, \epsilon_{i,t}, \mathcal{R}(\boldsymbol{x}_{i,t}, \epsilon_{i,t})\}$, where $t^s_{i,t} \leq t \leq t^e_{i,t}$.
Then $(t^e_{i,t} - t^s_{i,t})$ is the length of the interaction history and 
$\mathcal{Q}_{i} = \{t^s_{i,t} \leq t \leq t^e_{i,t},\,|\, q_{i,t} = 1, t \in \mathbb{N} \}$ is the set of time slots when there was a packet arrival. 
The environment descriptor can be estimated as follows:
$\tilde{\lambda}_{i,t} \approx |\mathcal{Q}_{i}|/(t^e_{i,t} - t^s_{i,t})$,
$\tilde{a}_{i,t} \approx  \sum_{t \in \mathcal{Q}_{i}} (b_{i,t+1} - b_{i,t})/ |\mathcal{Q}_{i}|$, and 
$\tilde{\sigma}^2_{i,t} \approx [\sum_{t \in \mathcal{Q}_{i}} (b_{i,t+1} - b_{i,t} - \bar{a}_{i,t})]/ |\mathcal{Q}_{i}| $.
% where $t \in \mathcal{Q}_{i}$.
Here, the device-specific parameters $\kappa_{i}$ and $\epsilon_{i,\textrm{max}}$ can be directly obtained from the IoT device. By combining the estimated values of $\tilde{\lambda}_{i,t}$, $\tilde{a}_{i,t}$ and $\tilde{\sigma}_{i,t}^2$, the environment descriptor $\tilde{\boldsymbol{z}}_{i,t}$ can be obtained. This process is dubbed as \emph{environment discovery}.

Up until now, there are two variables that capture these high-level features for each single environment: i) the environment policy $\boldsymbol{\theta}_{i,t}$ and ii) the environment feature vector $\phi(\boldsymbol{z}_{i,t})$. Here, $\boldsymbol{\theta}_{i,t}$ determines the cost of each device at any time slot, while $\phi(\boldsymbol{z}_{i,t})$ identifies a specific environment. Both variables are independently distributed, as the environment distribution is i.i.d and unknown. Thus, $\boldsymbol{\theta}_{i,t}$ and $\phi(\boldsymbol{z}_{i,t})$ contain different attributes of an environment~\cite{yu2014discriminative}. Hence, it is worth exploiting the similarities between these attributes to share experiences across various environments.
\vspace{-0.09cm}
\subsection{Lifelong Learning for Non-stationary Environments}

Our goal is to achieve a balance between the AoI and energy consumption for all the devices. Given that the devices experience independent environments and with the aforementioned MDP model, the problem \eqref{Problem_1_LV} can be rewritten as:  
\begin{equation}
\min_{ \Pi_{i,t}} \dfrac{1}{Z}    \sum_{i=1}^{N}  \sum_{t=0}^{T} \mathcal{J}(\boldsymbol{\theta}_{i,t}) \, ,
\label{J_theta}
\end{equation}
where
$\mathcal{J}(\boldsymbol{\theta}_{i,t}) = \int p_{\boldsymbol{\theta}_{i,t}}(\boldsymbol{\tau}) \mathfrak{R_{i,t}}(\boldsymbol{\tau}) \textrm{d} \boldsymbol{\tau}$, $p_{\boldsymbol{\theta}_{i,t}}$ is the probability distribution of interaction history $\boldsymbol{\tau}$, and $\mathfrak{R_{i,t}}(\boldsymbol{\tau})$ is the reward of the trajectory $\tau$.
As such, we can formulate the aforementioned variables as:
\begin{equation}
p_{\boldsymbol{\theta}_{i,t}}(\boldsymbol{\tau})\! =\! P_0(\boldsymbol{x}_0) \prod_{t = t^s_{i,t}}^{t^e_{i,t}}  p(\boldsymbol{x}_{i,t+1}| \boldsymbol{x}_{i,t}, \epsilon_{i,t}  )\, \pi_{{\boldsymbol{\theta}}_{i,t}}(\epsilon_{i,t} | \boldsymbol{x}_{i,t}),
\end{equation}
\begin{equation}
\mathfrak{R_{i,t}}(\boldsymbol{\tau}) = \dfrac{1}{t^s_{i,t} - t^e_{i,t}} \sum_{t = t^s_{i,t}}^{t^e_{i,t}}  \mathcal{R}(\boldsymbol{x}_{i,t}, \epsilon_{i,t}),
\end{equation}
where $p(\boldsymbol{x}_{i,t+1}| \boldsymbol{x}_{i,t}, \epsilon_{i,t}  )$ is the unknown state transition probability that maps a state-action pair at time slot $t$ onto a distribution of states at time slot $t+1$.
% In addition, $t^s$ and $t^e$ are the starting and ending time of the trajectory $\boldsymbol{\tau}$.

\begin{algorithm}
	\caption{UpdateL}
	\label{UpdateL}
	\begin{algorithmic}
		\IF{$\chi_{i} = 0$}
		\STATE $\boldsymbol{A}_{\boldsymbol{L}} \leftarrow \boldsymbol{A}_{\boldsymbol{L}} - \big( \boldsymbol{s} \boldsymbol{s}^\mathrm{T}  \big) \otimes \boldsymbol{\Gamma} $
		\STATE $\boldsymbol{b}_{\boldsymbol{L}} \leftarrow \boldsymbol{b}_{\boldsymbol{L}} - \mathrm{vec} \big( \boldsymbol{s}^\mathrm{T}  \otimes \big(   \boldsymbol{\alpha}^\mathrm{T} \boldsymbol{\Gamma}  \big) \big)$
		\ELSE
		\STATE Identify a new environment for device $i$
		\STATE $Z \leftarrow Z+1$
		\ENDIF
		\STATE $\boldsymbol{A}_{\boldsymbol{L}} \leftarrow \boldsymbol{A}_{\boldsymbol{L}} + \big( \boldsymbol{s} \boldsymbol{s}^\mathrm{T}  \big) \otimes \boldsymbol{\Gamma} $
		\STATE $\boldsymbol{b}_{\boldsymbol{L}} \leftarrow \boldsymbol{b}_{\boldsymbol{L}} + \mathrm{vec} \big( \boldsymbol{s}^\mathrm{T}  \otimes \big(   \boldsymbol{\alpha}^\mathrm{T} \boldsymbol{\Gamma}  \big) \big)$
		\STATE $\boldsymbol{L} \leftarrow \mathrm{mat} \Big(  \big( \frac{1}{Z} \boldsymbol{A}_{\boldsymbol{L}} + \eta_3 \boldsymbol{I}_{d \times h, d \times h} \big) ^{-1} \frac{1}{Z} \boldsymbol{b}_{\boldsymbol{L}} \Big) $
	\end{algorithmic}
\end{algorithm}
\begin{algorithm}
	\caption{Lifelong Reinforcement Learning}
	\label{PG-ELLAAlgo}
	\begin{algorithmic}
		\REQUIRE $T \leftarrow 0$, $\boldsymbol{A} \leftarrow \mathrm{zeros}_{d \times h, d \times h}$, $\boldsymbol{b} \leftarrow \mathrm{zeros}_{d \times h, 1}$
		\REQUIRE $\boldsymbol{L} \leftarrow \mathrm{zeros}_{d, h}$, $\boldsymbol{D} \leftarrow \mathrm{zeros}_{d_{\emph{z}}, h}$
		% \WHILE{some device $i$ is available}
		\WHILE{UAV arrives at device $i$}
		\STATE Collect $\boldsymbol{\tau}$ %\leftarrow \mathrm{Collect\,Interaction\, History}$
		\STATE Identify environment feature $\phi(\tilde{\boldsymbol{z}})$ for device $i$ using $\boldsymbol{\tau}$
		\STATE Update $\chi_{i}$ according to $\phi(\tilde{\boldsymbol{z}})$ 
		\STATE Compute $\boldsymbol{\alpha}_{i,t}$  and $\boldsymbol{\Gamma}_{i,t}$ from $\boldsymbol{\tau} $
		\STATE $\boldsymbol{L}, \boldsymbol{D} \leftarrow \mathrm{reinitializeAllZeroColumns}(\boldsymbol{L},\boldsymbol{D})$
		\STATE $\boldsymbol{s} \leftarrow \arg \min_{\boldsymbol{s}} \ell \big(  \boldsymbol{K},  \boldsymbol{s}, \boldsymbol{\beta}, \boldsymbol{Q} \big)$
		\STATE $\boldsymbol{L} \leftarrow \mathrm{updateL} \ell(\boldsymbol{L},  \boldsymbol{s}, \boldsymbol{\alpha}, \boldsymbol{\Gamma})$
		\STATE $\boldsymbol{D} \leftarrow \mathrm{updateD} \ell(\boldsymbol{D},  \boldsymbol{s}, \boldsymbol{\phi(\tilde{\emph{z}})}, \eta_1 \boldsymbol{I}_{d_z})$
		\ENDWHILE
	\end{algorithmic}
\end{algorithm}

The problem in \eqref{J_theta} involves a sequential stream of independent reinforcement environments as denoted in \ref{IIIA}.
% The problem in \eqref{J_theta} involves optimizing the performance of IoT devices in an environment that is constantly changing and unpredictable. 
% In order to address the uncertainties of the environments, we employ Lifelong Learning that develops intuitions for new environments. 
To enable continuous learning throughout the dynamical RL environments, it is important to explore and exploit the common structures revealed by the high level features such as $\boldsymbol{\theta}_{i,t}$ and $\phi(\boldsymbol{z}_{i,t})$.
% To exploit the commonalities across environments, the MDP structures comes naturally.
% To do so, we assume that the task feature vector can be a linear combination of $h$ latent components such that $\phi(\emph{z}_{i,j}) = \boldsymbol{D}\boldsymbol{s}_{i,j}$, where $\boldsymbol{s}_{i,j} \in \mathbb{R}^{h}$ is a vector of linear parameters and $\boldsymbol{D}$ is a knowledge base with a library of $h$ latent components. The dimension of $h$ is chosen independently via cross-validation.
% Because the task feature basis can capture the commonalities among task descriptors, these kinds of commonalities can be reflected into the sparse library $\boldsymbol{D}$ and $\boldsymbol{s}_{i,j}$.
% l 和 d 这两段，可不可以也提炼一下共同的内容。比如，sparse的部分就是相同的。
To enable knowledge transfer between environments, we assume that the policy $\boldsymbol{\theta}_{i,t}$ is a linear combination of $h$ latent components \cite{kumar2012learning}, i.e., $\boldsymbol{\theta}_{i,t} = \boldsymbol{L}\boldsymbol{s}_{i,t}$, where $\boldsymbol{s}_{i,t} \in \mathbb{R}^{h}$ is a vector of linear parameters and $\boldsymbol{L}$ is a knowledge base with a library of $h$ latent components that represents the shared knowledge of all the environments.
The dimension of $h$ is chosen independently with cross-validation. The mapping function $\boldsymbol{s}_{i,t}$ should be sparse to maximize the knowledge captured by the latent components.
As such, each observed environment can be a linear combination of only a few latent components in $\boldsymbol{L}$. 
To incorporate the environment feature, we also assume that the environment feature vector can be linearly represented by a latent basis $\boldsymbol{D} \in \mathbb{R}^{d_z \times h}$.
Similar to the knowledge base $\boldsymbol{L}$, the environment feature basis $\boldsymbol{D}$ can capture the commonalities among the environment descriptors, such as $\phi(\boldsymbol{z}_{i,t}) =  \boldsymbol{D} \boldsymbol{s}_{i,t}$.
As illustrated in Fig.~\ref{LSD}, the policy base $\boldsymbol{L}$ and feature base $\boldsymbol{D}$ share the same coefficient vectors $\boldsymbol{s}_{i,t}$.
As such, the environment policy $\boldsymbol{\theta}_{i,t}$ and the environment feature vector $\phi(\boldsymbol{z}_{i,t})$ can be connected through the shared mapping vectors.
It is reasonable to utilize the relationship between the feature dictionary and the knowledge base.
% Next, we will show that the feature dictionary and the knowledge base will augment each other through the same learning process. 

%  $\phi(\emph{z}_{i,j}) =  \boldsymbol{D} \boldsymbol{s}_{i,j}$ and 
\begin{figure}
	\centerline
	{\includegraphics[width=3.5 in]{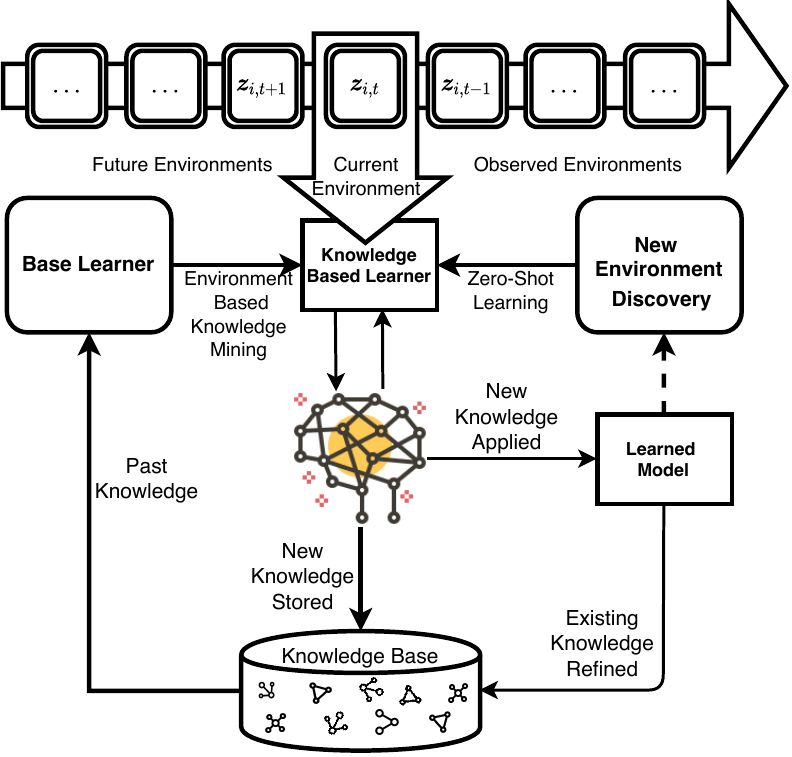}}
	\caption{\small The flow diagram of the proposed lifelong RL algorithm.}
	\label{Flow}
\end{figure}

We proceed to optimize the coupled bases $\boldsymbol{L}$ and $\boldsymbol{D}$ together. Up to this end, techniques from the field of sparse coding are utilized. Specifically, coupled dictionary optimization \cite{yang2010image} is applied to optimize the dictionaries for multiple feature spaces with a joint sparse representation. The result of incorporating knowledge transfer and feature coding into the optimization process is a multi-environment loss function based on coupled dictionaries, given as:
\begin{alignat}{2}
g_\textrm{T} (\boldsymbol{L}, \boldsymbol{D})  =
& \dfrac{1}{Z} \sum_{i = 1}^{N} \sum_{t = 0}^{T} \min_{\boldsymbol{s}_{i,t}} 
\Big[ 
\mathcal{J}(\boldsymbol{\theta}_{i,t}) 
+ \eta_1 \lVert \phi(\boldsymbol{z}_{i,t}) - \boldsymbol{D} \boldsymbol{s}_{i,t}  \lVert_{2}^2  \nonumber \\
& + \eta_2  \lVert \boldsymbol{s}_{i,t}  \lVert_{1} 
\Big]
+ \eta_3 (\lVert \boldsymbol{L}  \lVert_{\textrm{F}}^2  + \lVert \boldsymbol{D}  \lVert_{\textrm{F}}^2)\, ,
\label{Target_1}
\end{alignat}
where $\ell_1$-norm approximates the vector sparsity and $\lVert \boldsymbol{L} \lVert_{\textrm{F}} = (tr(\boldsymbol{L}\boldsymbol{L}'))^{1/2}$ is the Frobenius norm of matrix $\boldsymbol{L}$. The parameter $\eta_1$ controls the balance between the policy's fit and the feature's fit. Also, $\eta_2$ and $\eta_3$ are two regularization parameters, where $\eta_2$ controls the sparsity of $\boldsymbol{s}_{i,t}$. The penalty on the Frobenius norm of $\boldsymbol{L}$ and $\boldsymbol{D}$ regularizes the predictor weights to have low $\ell_2$-norm and avoids overfitting.

The optimal $\boldsymbol{\theta}_{i,t}$ in \eqref{Target_1} can achieve the minimum $g_\textrm{T} (\boldsymbol{L}, \boldsymbol{D})$, having $\boldsymbol{L}$ and $\boldsymbol{D}$ given. 
Considering that $\boldsymbol{\theta}_{i,t} = \boldsymbol{L} \boldsymbol{s}_{i,t}$, \eqref{Target_1} is transformed into a minimization problem based on $\{\boldsymbol{s}_{i,t}\}$.
% In \eqref{Target_1} and for any $i$, the optimal $\boldsymbol{\theta}_{i,t}$ can achieve the minimum $g_\textrm{T} (\boldsymbol{L}, \boldsymbol{D})$ with given $\boldsymbol{L}$ and $\boldsymbol{D}$. 
% Considering $\boldsymbol{\theta}_{i,t} = \boldsymbol{L} \boldsymbol{s}_{i,t}$, \eqref{Target_1} is transformed into a minimization problem based on $\{\boldsymbol{s}_{i,t}\}$.
% With the assumption that the knowledge bases $\boldsymbol{L}$ and $\boldsymbol{D}$ is available, the $g_\textrm{T} (\boldsymbol{L}, \boldsymbol{D})$ can be minimized when all the tasks $\phi(\emph{z}_{i,j})$ are available. 
% However, it is easier to get L and D with all the tasks available, i.e., the $\phi(\emph{z}_{i,j})$. 
% The ideal case is that all the tasks $\emph{z}_{i,j}$ are available. When the base L and D and the feature vector is available, the optimal policy theta can be obtained. However, 
% It is obvious that \eqref{Target_1} is ideal situation in which all the tasks $\{\emph{z}_{i,j}\}$ are available. 
% With the assumption that the knowledge bases $\boldsymbol{L}$ and $\boldsymbol{D}$ is available, the $g_\textrm{T} (\boldsymbol{L}, \boldsymbol{D})$ can be minimized when all the tasks $\emph{z}_{i,j}$ are available. 
% With all the optimal $\theta_{i,j}$ and the relationship between $\theta$ and $s_{i,j}$, the equation is transformed into a minimization problem based on $s_{i,j}$. 
% As such, with the base L and D available, the minimal gT can be obtained either. In this case, the optimal $s_{i,j}$ and $\theta$ will be obtained accordingly.
Consequently, we can first obtain $\boldsymbol{L}$ and $\boldsymbol{D}$ with a series of consecutive environments. Then, $\boldsymbol{s}_{i,t}$ can be further optimized. Given $\boldsymbol{L}$ and $\boldsymbol{s}_{i,t}$, the suboptimal $\boldsymbol{\theta}_{i,t}$ can be obtained eventually.

To compute $\boldsymbol{L}$ and $\boldsymbol{D}$, we need to access the interaction history of all environments for all devices, as evident from \eqref{Target_1}. 
Notably, there remains dependence between the policies of environments and their interaction history. To suppress this dependence, we use a second-order Taylor expansion to approximate $\mathcal{J}(\boldsymbol{\theta}_{i,t})$ around an estimated optimal policy, denoted as $\boldsymbol{\alpha}_{i,t}$, for each individual environment. 
% This estimated policy is the optimal policy for each environment. 
The estimated policy is defined as:
\begin{equation}
\boldsymbol{\alpha}_{i,t} = \arg \min_{\boldsymbol{\theta}_{i,t}} \mathcal{J}(\boldsymbol{\theta}_{i,t}) .
% + \eta_{s}  \lVert \boldsymbol{\theta}_{i,j}  \lVert_{2}^2 
\label{alpha}
\end{equation}
% where $\eta_{s}$ is a regularization parameter. 
Here, the method used to obtain $\boldsymbol{\alpha}_{i,t}$ is the base learner. Accordingly, we use the collected interaction history of device $i$ to evaluate $\boldsymbol{\alpha}_{i,t}$.

The second-order Taylor expression of $\mathcal{J}(\boldsymbol{\theta}_{i,t})$ is expanded around $\boldsymbol{\alpha}_{i,t}$. For each environment $\emph{z}_{i,t}$, 
\begin{alignat}{2}
\mathcal{J}(\boldsymbol{\theta}_{i,t} = \boldsymbol{L}\boldsymbol{s}_{i,t}) 
=  &
\mathcal{J}(\boldsymbol{\alpha}_{i,t})
+ \nabla \mathcal{J}(\boldsymbol{\theta}_{i,t})_{\boldsymbol{\theta}_{i,t} = \boldsymbol{\alpha}_{i,t}} (\boldsymbol{\alpha}_{i,t} -\boldsymbol{L}\boldsymbol{s}_{i,t} ) \nonumber \\
& + \lVert \boldsymbol{\alpha}_{i,t} -\boldsymbol{L}\boldsymbol{s}_{i,t} \lVert^2_{\boldsymbol{\Gamma}_{i,t}} ,
\label{second-order}
\end{alignat}
where $\nabla \mathcal{J}(\boldsymbol{\theta}_{i,t})$ is the first-order gradient of $\mathcal{J}(\boldsymbol{\theta}_{i,t})$ and $\boldsymbol{\Gamma}_{i,t}$ is the Hessian matrix. % \cite{ammar2014online}
The first term on the right-hand side (RHS) of \eqref{second-order} is a constant and can be suppressed.
The second term takes a negligible value as $\boldsymbol{\alpha}_{i,t}$ is the minimizer of $\eqref{alpha}$.
Substituting the second-order Taylor expansion into  \eqref{Target_1} yields the following loss function:
\begin{alignat}{2}
g_\textrm{T} (\boldsymbol{L}, \boldsymbol{D})  =
& \dfrac{1}{Z} \sum_{i = 1}^{N} \sum_{t = 0}^{T} \min_{\boldsymbol{s}_{i,t}} 
\Big[ 
\lVert \boldsymbol{\alpha}_{i,t} -\boldsymbol{L}\boldsymbol{s}_{i,t} \lVert^2_{\boldsymbol{\Gamma}_{i,t}} \nonumber \\
&+ \eta_1 \lVert \phi(\boldsymbol{z}_{i,t}) - \boldsymbol{D} \boldsymbol{s}_{i,t}  \lVert_{2}^2  
 + \eta_2  \lVert \boldsymbol{s}_{i,t}  \lVert_{1} 
\Big] \nonumber \\
& + \eta_3 (\lVert \boldsymbol{L}  \lVert_{\textrm{F}}^2  + \lVert \boldsymbol{D}  \lVert_{\textrm{F}}^2)\, .
\label{Target_2}
\end{alignat}
\begin{algorithm}[t]
	\caption{Zero-Shot Transfer for a New Task}
	\label{zeroshotnewtask}
	\begin{algorithmic}
		\REQUIRE Trained $\boldsymbol{L}  \in \mathbb{R}^{d \times h}$ and $\boldsymbol{D}  \in \mathbb{R}^{d_z \times h}$
		\STATE Estimate $\phi(\tilde{\emph{z}}_{i,t})$ from collected interaction history $\tau$
		\STATE $\boldsymbol{s}^*_{i,t} \leftarrow \arg \min_{\boldsymbol{s}_{i,t}} \lVert \phi(\tilde{\boldsymbol{z}}_{i,t}) - \boldsymbol{D} \boldsymbol{s}_{i,t} \lVert_{2}^2 + \eta_2 \lVert \boldsymbol{s}_{i,t} \lVert_{1}$
		\STATE $\boldsymbol{\theta}^*_{i,t} = \boldsymbol{L} \boldsymbol{s}^*_{i,t}$
	\end{algorithmic}
\end{algorithm} 
Now, considering the symmetrical characteristic in \eqref{Target_2}, the following pairs can be abstracted as follows:
\begin{equation}
\boldsymbol{\beta}_{i,t} = \bigg[ 
\begin{matrix}
\boldsymbol{\alpha}_{i,t} \\
\phi(\boldsymbol{z}_{i,t}) 
\end{matrix}
\bigg];
\quad
\boldsymbol{K} = \bigg[ 
\begin{matrix}
\boldsymbol{L} \\
\boldsymbol{D} 
\end{matrix}
\bigg];
\quad
\boldsymbol{Q}_{i,t} = 
\bigg[ 
\begin{matrix}
\boldsymbol{\Gamma}_{i,t} & \boldsymbol{0} \\
\boldsymbol{0} & \eta_1 \boldsymbol{I}_{d_z}
\end{matrix}
\bigg] \; ,
\end{equation}
where $\boldsymbol{0}$ is the all-zero matrix. With this abstraction, \eqref{Target_2} is simplified to the following:
\begin{alignat}{2}
g_\textrm{T} (\boldsymbol{K})  =
& \dfrac{1}{Z} \sum_{i = 1}^{N} \sum_{t = 0}^{T} \min_{\boldsymbol{s}_{i,t}} 
\Big[ 
\lVert \boldsymbol{\beta}_{i,t} -\boldsymbol{K}\boldsymbol{s}_{i,t} \lVert^2_{\boldsymbol{Q}_{i,t}} \nonumber \\
&+ \eta_2  \lVert \boldsymbol{s}_{i,t}  \lVert_{1} 
\Big] 
+ \eta_3 \lVert \boldsymbol{K}  \lVert_{\textrm{F}}^2 \;  .
\label{Target_3}
\end{alignat}

% \subsubsection{Dependence Suppression}
 
% Batch MTL is enabled by \eqref{Target_3}, but online MTL is not possible due to two reasons: (1) it requires a series of interaction histories $\boldsymbol{\tau}$ for all devices in all environments, and (2) optimizing $\boldsymbol{L}$ and $\boldsymbol{D}$ requires the evaluation of $\boldsymbol{s}_{i,j}$ through all values, which is computationally intensive. Therefore, direct MTL optimization is impossible, and we use an alternating optimization method instead.

Clearly, \eqref{Target_3} is a joint optimization problem for $\boldsymbol{\beta}_{i,t}$ and $\boldsymbol{K}$. Solving this problem requires the availability of trajectories of all the devices throughout the time period $T$ to compute the Hessian matrices $\boldsymbol{\Gamma}_{i,t}$, i.e., $\boldsymbol{Q}_{i,t}$.
However, this is not feasible in practice. Firstly, predicting future dynamics beforehand is unattainable. Secondly, the UAV is limited to accessing one device at any given moment. Hence, only the trajectory of the currently visited device by the UAV can be obtained.
Thus, an approach that can eliminate these dependencies is necessary.

Next, we adopt an alternating approach to optimize the shared knowledge base and device specific knowledge.
First, we leverage the initialized $\boldsymbol{K}$ to optimize $\boldsymbol{s}_{i,t}$ of the current device and the environment it is experiencing. Then, we utilize the obtained $\boldsymbol{s}_{i,t}$ to optimize $\boldsymbol{L}$ and $\boldsymbol{D}$.
% We propose to optimize $\boldsymbol{s}_{i,t}$ only when the UAV is visiting device $i$, and update $\boldsymbol{L}$ and $\boldsymbol{D}$ each time $\boldsymbol{s}_{i,t}$ is updated. 
% To solve for the unknown $\boldsymbol{K}$ and $\boldsymbol{s}_{i,t}$ in the loss function \eqref{loss}, we first use the initialized $\boldsymbol{K}$ to optimize $\boldsymbol{s}_{i,t}$ as in \eqref{s}, and then use the obtained $\boldsymbol{s}_{i,t}$ to optimize $\boldsymbol{L}$ and $\boldsymbol{D}$ as in \eqref{L_update}. To eliminate the dependence and apply this method in our online setting, we adopt alternating optimization and update each device's $\boldsymbol{s}_{i,t}$ using \eqref{Target_2} to suppress the dependence on the other devices.
% Accordingly, we define the following loss function:
With that in mind, we can rewrite \eqref{Target_3} as follows:
\begin{equation}
	\boldsymbol{s}_{i,t} \leftarrow \arg \min_{\boldsymbol{s}_{i,t}} \ell \big(  \boldsymbol{K},  \boldsymbol{s}_{i,t}, \boldsymbol{\beta}_{i,t}, \boldsymbol{Q}_{i,t} \big)  \, ,
	\label{s} 
\end{equation}
\begin{equation}
	\boldsymbol{K} = \arg \min_{\boldsymbol{K}} \dfrac{1}{Z} \sum_{i=1}^{N} \sum_{t=0}^{T} \ell \big(  \boldsymbol{\boldsymbol{K}},  \boldsymbol{s}_{i,t}, \boldsymbol{\beta}_{i,t}, \boldsymbol{Q}_{i,t} \big)
	+
	\eta_3 \lVert \boldsymbol{\boldsymbol{K}}  \lVert_{\textrm{F}}^2 \, ,
	\label{K_update}
\end{equation}
where
\begin{equation}
	\ell \big(  \boldsymbol{K},  \boldsymbol{s}_{i,t}, \boldsymbol{\beta}_{i,t}, \boldsymbol{Q}_{i,t} \big) =
	\lVert  \boldsymbol{\beta}_{i,t} - \boldsymbol{K} \boldsymbol{s}_{i,t}  \lVert^2_{\boldsymbol{Q}_{i,t}}
	+ \eta_2  \lVert \boldsymbol{s}_{i,t}  \lVert_{1}  \, .
	\label{loss}
\end{equation}
By fixing $\boldsymbol{K}$, $\boldsymbol{s}_{i,t}$ can be updated as in \eqref{s}. This is an $\ell_1$-regularized regression problem that can be solved as an instance of Lasso.

To update $\boldsymbol{K}$, we decouple \eqref{K_update} between $\boldsymbol{L}$ and $\boldsymbol{D}$. Since the two variables have similar structures, their update processes are identical. Hence, we consider $\boldsymbol{L}$ as an example and define its associated loss function $\ell \big( \boldsymbol{\boldsymbol{L}}, \boldsymbol{s}_{i,t}, \boldsymbol{\alpha}_{i,t}, \boldsymbol{\Gamma}_{i,t} \big)$.
Accordingly, $\boldsymbol{L}$ is updated through:
\begin{equation}
	\boldsymbol{L} = \arg \min_{\boldsymbol{L}} \dfrac{1}{Z} \sum_{i=1}^{N} \sum_{t=0}^{T} \ell \big(  \boldsymbol{\boldsymbol{L}},  \boldsymbol{s}_{i,t}, \boldsymbol{\alpha}_{i,t}, \boldsymbol{\Gamma}_{i,t} \big)
	+
	\eta_3 \lVert \boldsymbol{\boldsymbol{L}}  \lVert_{\textrm{F}}^2 \, .
	\label{L_update}
\end{equation}
Herein, $\boldsymbol{L}$ can be easily obtained by nulling the gradient of \eqref{L_update}. Then $\boldsymbol{L}$ can be obtained as $\boldsymbol{A}^{-1}_{\boldsymbol{L}}\boldsymbol{b}_{\boldsymbol{L}}$, where:
\begin{equation}
\boldsymbol{A}_{\boldsymbol{L}} = \eta_3 \boldsymbol{I}_{d \times h, d \times h} + \dfrac{1}{Z} \sum_{i=1}^{N} \sum_{t=0}^{T}
\big( \boldsymbol{s}_{i,t} \boldsymbol{s}_{i,t}^\mathrm{T}  \big)
\otimes \boldsymbol{\Gamma}_{i,t} ,
\label{A}
\end{equation}
\begin{equation}
\boldsymbol{b}_{\boldsymbol{L}} = \dfrac{1}{Z} \sum_{i=1}^{N} \sum_{t=0}^{T}  \mathrm{vec}
\big( \boldsymbol{s}_{i,t}^\mathrm{T}  \otimes \big(   \boldsymbol{\alpha}_{i,t}^\mathrm{T} \boldsymbol{\Gamma}_{i,t}  \big) \big) \, .
\label{b}
\end{equation}
The UAV repeats the above process for each environment until $\boldsymbol{L}$
%  and $\boldsymbol{D}$ 
converges.
This process is referred to as \emph{updateL}, that is summarized in Algorithm \ref{UpdateL}.
% The full steps can be found in Algorithm \ref{UpdateL}. $\boldsymbol{s}$, $\boldsymbol{\alpha}$ and $\boldsymbol{\Gamma}$ provide the assemble set over $i$ and $j$. A similar update algorithm can be used to update $\boldsymbol{D}$.
% By updating $\boldsymbol{L}$, $\boldsymbol{D}$, and $\boldsymbol{s}_{i,j}$ alternately, {\color{red} the dependencies we mentioned above are eliminated.} 
% Though only one device is considered each time, the policy improvement of other devices can be obtained through the improvement of the knowledge basis $\boldsymbol{L}$ and feature dictionary $\boldsymbol{D}$.
% The complete flow of the approach is presented in Algorithm \ref{PG-ELLAAlgo}.
% The corresponding flow diagram can be found in Fig.~\ref{Flow}.
% After proper training, the UAV can quickly find the sub-optimal policy for each environment it encountered with this approach.
% The above algorithm requires access to the optimal policy $\boldsymbol{\alpha}_{i,j}$.
% To compute it, however, a large number of iterations are required. 
% Each iteration requires a set of interaction history, making the training process computation extensive and time consuming.
% More importantly, it cannot handle the new environment that the UAV has never encountered.
% To address these issues, a zero-shot method is introduced in the next subsection.
% Herein, algorithm \ref{UpdateL} updates $\boldsymbol{L}$ and $\boldsymbol{s}_{i,t}$. 
By assembling $\boldsymbol{s}_{i,t}$, $\boldsymbol{\alpha}_{i,t}$, and $\boldsymbol{\Gamma}$ over $i$ and $t$, a similar algorithm can be used to update $\boldsymbol{D}$, which is referred to as \emph{updateD}.
% in the following. Through this approach, 
Even though only a single environment is considered at a time, the policy improvement of the other environments can be obtained by improving the knowledge base $\boldsymbol{L}$ and the feature dictionary $\boldsymbol{D}$. The complete flow of this approach is presented in Algorithm \ref{PG-ELLAAlgo}, and a corresponding flow diagram is presented in Fig.~\ref{Flow} to summarize this approach. After proper training, the UAV can quickly find a high-fidelity policy $\mathcal{E}^* = \{\epsilon^*_{i,t}\}$ for each environment encountered.

\begin{figure}
	\centering
	\includegraphics[width=3.5in]{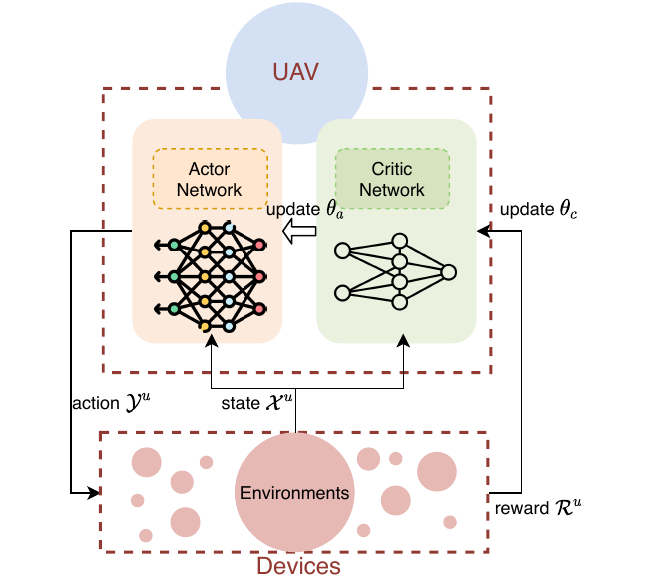}
	\caption{\small An illustration of the proposed AC network for UAV flight control.}
	\label{figAC}
\end{figure}

\subsection{Zero-Shot Learning Method}

Thus far, we have achieved the learning and knowledge accumulation for environments with the help of $\boldsymbol{\alpha}_{i,t}$.
However, our objective is acquiring an algorithm that can adapt to unknown environments.
To further accelerate the learning of new environments, we adopt a zero-shot transfer method \cite{socher2013zero} with coupled dictionaries~\cite{romera2015embarrassingly}. 
The zero-shot method is often used to establish a connection between unobserved and observed classes in machine learning. 
% This is a method inspired by how humans learn new knowledge by reading its description and utilizing previously acquired concepts.
Thus, it is useful in the following cases: 
\begin{enumerate*}
  \item The unobserved classes are rare, and it is not easy to find adequate instances for training; 
  \item The total number of classes is large, and it is impossible to get all instances labeled or train all classes;
  \item It is expensive to obtain instances for certain classes;
  \item Target classes change over time and it is costly to label every class that is observed.
\end{enumerate*}

%Our situation is the combination of the above cases.
In the considered non-stationary environment, the new environments are not encountered before during during training. In other words, we do not have $\boldsymbol{\alpha}_{i,t}$ for these new environments.
Hence, we leverage the zero-shot method to associate the trained environments and new environments via the shared knowledge bases $\boldsymbol{L}$ and $\boldsymbol{D}$. 
Here, the feature descriptor $\phi(\boldsymbol{z}_{i,t})$ acts as a high-level descriptor of the learning process.

According to \cite{yang2010image}, a policy $\boldsymbol{\theta}_{i,t}$ in the policy parameter space can be recovered using the coupled dictionary $\boldsymbol{L}$ and $\boldsymbol{D}$ with a feature descriptor for each environment. We use the estimated value $\tilde{\boldsymbol{z}}_{i,t}$ obtained with a small set of interaction histories $\boldsymbol{\tau}$, and then obtain the shared mapping vector using the estimated feature descriptor $\phi(\tilde{\boldsymbol{z}}_{i,t})$.
The loss function $\ell(\boldsymbol{D}, \boldsymbol{s}, \boldsymbol{\phi(\tilde{\boldsymbol{z}})}, \eta_1 \boldsymbol{I}_{d_z})$ in \eqref{loss} can be rewritten as: 
\begin{equation}
\boldsymbol{s}^*_{i,t} \leftarrow \arg \min_{\boldsymbol{s}_{i,t}} \{\lVert \phi(\tilde{\boldsymbol{z}}_{i,t}) - \boldsymbol{D} \boldsymbol{s}_{i,t} \lVert_{2}^2 + \eta_2 \lVert \boldsymbol{s}_{i,t} \lVert_{1} \} .
\label{zero-shot}
\end{equation}
% The policy parameter $\boldsymbol{\theta}^*_{i,j}$ can further be obtained as $\boldsymbol{\theta}^*_{i,j} = \boldsymbol{L} \boldsymbol{s}^*_{i,j}$. 
% The sub-optimal interaction decision $\epsilon^*_{i,j,t}$ can be obtained through a stochastic policy, such as $  \epsilon^*_{i,j,t} = \boldsymbol{\theta}^*_{i,j}\boldsymbol{x}_{i,j,t} \, +  \, c$, with $c\thicksim \mathcal{N} (0, \,  c^2)$.
% A detailed procedure can be found in Algorithm 3.
% Now, we are able to extend Algorithm \eqref{PG-ELLAAlgo} to brandly new environments with minimum time delay and small computation consumption.
Having obtained the policy parameter $\boldsymbol{\theta}^*_{i,t}$, the optimal interacting decision $\mathcal{E}^* = \{\epsilon^*_{i,t}\}$ can be determined using a stochastic policy, i.e., $ \epsilon^*_{i,t} = \boldsymbol{\theta}^*_{i,t}\boldsymbol{x}_{i,t}  +  \epsilon_z $, where $\epsilon_z$ is the noise of the stochastic policy having
% $\sigma'$ follows a Gaussian distribution with mean 0 and variation .
$\epsilon_z \thicksim \mathcal{N} (0, \sigma_z^2)$ with $\sigma_z$ being the standard deviation of Gaussian distribution. 
This procedure is summarized in Algorithm~\ref{zeroshotnewtask}.

\section{Energy Efficient UAV Trajectory and Velocity Optimization}\label{sec: UAV energy}

% In this part, we will study the UAV's flying decisions $\mathcal{L} = \{\boldsymbol{l}_m\}$ and $\mathcal{V} = \{v_m\}$ based on previously obtained $\mathcal{E}^* = \{\epsilon^*_{i,j,t}\}$.
% In this case, UAV not only serves as a central agent who is responsible for maintaining knowledge bases $\boldsymbol{L}$ and $\boldsymbol{D}$ for all the devices, but also needs to make flying decisions of itself through out the time.
% In the below, we will focus on the following objective function:

In this section, we optimize the flight decisions $\boldsymbol{F}$ and $\boldsymbol{v}$, i.e., trajectory and velocity of the UAV, respectively.  
% Specifically, we study the variables, 
% i.e., 
% and how they relate to the UAV's decision-making process. 
% The UAV operates in a complex environment. 
We note here that the UAV is responsible for maintaining the knowledge bases $\boldsymbol{L}$ and $\boldsymbol{D}$ for all devices and taking the corresponding flight decisions.

The flight control objective of the UAV can be formulated in the form of an optimization problem as follows:
\begin{alignat}{2}
	\min_{  \boldsymbol{v}  , \, \boldsymbol{F} } \quad & \mu \dfrac{1}{M}  \sum_{m=0}^{M}  
	e^U_m(v_m, \boldsymbol{l}_m, \boldsymbol{l}_{m+1})+ \frac{1}{Z} \sum_{i=1}^N \sum_{t=0}^{T} c^*_{i}(t)  \label{Target_UAV} \\      
	\mathrm{s.t.} \quad 
	%& \lim_{M\to\infty}\dfrac{1}{M}  \sum_{m=0}^{M} e^u_m \leq \bar{e}^u  \nonumber \, ,\\
	& \eqref{eqn3}, 
	% \eqref{eqn4}, 
	\eqref{eqn5} \, , \nonumber \\
	% & v_{\textrm{min}} \le v_m \le v_{\textrm{max}},  \quad \forall \, m= 0, 1, \ldots, M \, ,  \nonumber \\ 
	% & \boldsymbol{l}_m \in \{(x_i, y_i)\},   \quad \forall i= 1, 2, \ldots, N    \, , \nonumber \\
	& c^*_{i}(t) =  \beta \Delta_{i,t} + (1 - \beta) \kappa_i \epsilon_{i,t}^{*3} \,  . 
	% & \epsilon^*_{i,j,t} = \boldsymbol{\theta}^*_{i,j}\boldsymbol{x}_{i,j,t} \, +  \, c , \,  \textrm{Condition A}, \nonumber   \\
	% & \epsilon_{i,j,t}^{*} \in [0, \epsilon_{i,\textrm{max}}]  \,  , \textrm{Condition B} , \nonumber  
\end{alignat}

 Indeed, traditional RL may not be adequate to handle the non-stationarity surrounding the IoT devices in problem \eqref{Target_UAV}. However, it can proficiently manage flight control of the UAV over all the devices. This is due to the fact that the non-stationarity affects the distribution of the periods, i.e., $\bar{p}_i$ and $\sigma'_i$, of the IoT devices only. 
%This is due to the fact that the distribution of the periods of the non-stationary environments, i.e., $\bar{p}_i$ and $\sigma'_i$, varies across devices, however remains consistent for each device. 
As such, the flight decisions of the UAV are not affected the distribution changes in the environment of the devices, but rather by the frequency of change in the underlying environment. Thereby, the UAV operates at a higher level of abstraction from the IoT devices. 
%\textcolor{red}{Despite the existence of consistent patterns, these patterns remain undisclosed and the transition probabilities are likewise unknown.}
% From the perspective of the UAV, it is natural to employ the RL methods to capture the underlying regulations among devices and achieve optimal flight control.

Accordingly, RL methods such as value-based learning and policy-based learning can be potential solutions for UAV flight control. However, the unstable reward caused by the unpredictable environmental changes can slow down value-based learning, and the batch learning required for policy-based learning can greatly drain training resources.
Hence, we resort to an AC framework that offers increased stability, better convergence, and reduced variations. The AC network consists of two networks: i) the actor network that is responsible for decision-making and outputs actions based on the current state inputs, and ii) the critic network which interacts with the environment using the actions from the actor network and updates its value output accordingly. This value can be used as a judgment value in the actor network to increase or decrease the probability of the chosen action. The two networks interact with each other and the environment in an interative manner until the optimal flight policy is obtained. An illustration of the AC network functionality is described in Fig.~\ref{figAC}. Furthermore, we describe the denoted RL environment as follows:

\begin{itemize}
    \item 
The state space of the UAV, denoted as $\mathcal{X}^u \in \mathbb{R}^{N+2}$, includes the UAV's current location and the time that has elapsed since the current environment arrived at each device. Specifically, $\mathcal{X}^u = \{\varsigma_m \} = \{\boldsymbol{l}_{m}, \varpi_0, \varpi_1, \ldots , \varpi_N \}$.
If the UAV is visiting a device, the flight time for the current environment is added to $\varpi_i$. If it is a new environment, $\varpi_i$ is initialized according to the duration of the current environment. Accordingly, the corresponding flight time for the other devices increases.
    \item 
The action space of the UAV, denoted by $\mathcal{Y}^u$, is defined as its next destination $\boldsymbol{l}_{m+1}$ and velocity $\boldsymbol{v}_m$, i.e., $\mathcal{Y}^u = \{\boldsymbol{\zeta}_m\} = \{\boldsymbol{v}_m, \boldsymbol{l}_{m+1} | \boldsymbol{v}_m \le v_{\textrm{max}}, \boldsymbol{l}_{m+1} \in \mathbb{R}^2 \}$, since the UAV only accesses one device at a time.
    \item 
The reward of the UAV, denoted by $R^u(\varsigma_m, \boldsymbol{\zeta}m)$, is a function of the state and action; i.e., $R^u(\varsigma_m, \boldsymbol{\zeta}_m) = \mu e^U_m(\boldsymbol{v}_m, \boldsymbol{l}_m, \boldsymbol{l}_{m+1}) + \frac{1}{N} \sum_{i=1}^N \sum_{t_m}^{t=t_{m+1}} c_{i}(t)$. Here, each time the UAV selects a device to visit, it only considers the reward of that specific device in the current flight period.    
\end{itemize}

We use $\Pi^u_a = \{ \pi_{\boldsymbol{\theta}a} \}$ to denote the distribution of the actions over the states, and $\Pi^u_c = \{ \pi_{\boldsymbol{\theta}_c} \}$ to denote the action-value function parameter, where $\pi_{\boldsymbol{\theta}_a} (\boldsymbol{\zeta}_m|\varsigma_m) = \Pr\{\boldsymbol{\zeta}_m|\varsigma_m,\boldsymbol{\theta}_a \}$. 
\begin{figure}
	\centering
	\includegraphics[width=3.5in]{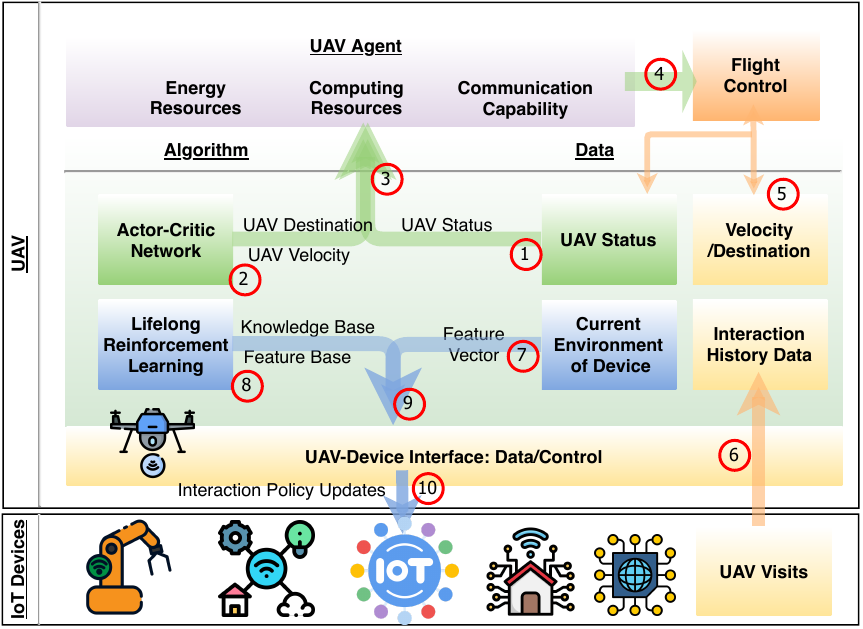}
	\caption{\small Illustrative figure of the proposed lifelong RL and AC network solution for the UAV-aided non-stationary IoT network optimization.}
	\label{figCom}
\end{figure}
The AC network maximizes the accumulated discounted reward $G_m =  -\sum_{k = m}^{\infty }  \eta_4^{k-m}  R^u(\varsigma_{k}, \boldsymbol{\zeta}_k)$, where $\eta_4 \in (0, 1]$ is the discount coefficient.
Given a state and the policy, the gain of an action is evaluated by an action-value function, i.e., the Q-function, that is denoted as:
\begin{equation}
	Q_{\pi_c} (\varsigma_{m}, \boldsymbol{\zeta}_m) = \mathbb{E}_{\boldsymbol{\theta}_c} [G_m |\varsigma_{m}, \boldsymbol{\zeta}_m  ]  \, .
\end{equation}
Furthermore, the Bellman expectation equation of the action-value function is given as: \begin{alignat}{2}
	Q_{\pi_c} (\varsigma_{m}, \boldsymbol{\zeta}_m) 
	= &\mathbb{E}_{R,\varsigma \thicksim \varXi  } 
	\bigl[
	R^u(\varsigma_m, \boldsymbol{\zeta}_m) \\ \nonumber
	& + \eta_4 \mathbb{E}_{\boldsymbol{\zeta}_{m+1}\thicksim \pi_a}[Q_{\pi_c} (\varsigma_{m+1}, \boldsymbol{\zeta}_{m+1})]\bigr] \, , 
\end{alignat}
where $\varXi$ is the non-statioinary environment that the devices interact with.
Thus, the actor network adjusts its policy based on the action-value function that is formulated as:
\begin{equation}
	\bigtriangledown_{\boldsymbol{\theta}_a} J(\boldsymbol{\theta}_a) 
	= \mathbb{E} 
	\bigg[
	\sum_{m = 0}^{\infty} 
	\bigtriangledown_{\boldsymbol{\theta}_a}
	\mathrm{log\pi_{\boldsymbol{\theta}_a}(\boldsymbol{\zeta}_m|\varsigma_m)
	Q_{\pi^c} (\varsigma_{m}, \boldsymbol{\zeta}_m) }
	\bigg] \, . 
\end{equation}
Then, the parameter of the actor network is updated through the following equation:
\begin{equation}
	\boldsymbol{\theta}_{a} \leftarrow  \boldsymbol{\theta}_{a} + \eta_a \bigtriangledown_{\boldsymbol{\theta}_a} J(\boldsymbol{\theta}_a)  \,  \label{eqAC},
\end{equation}
where $\eta_a$ is the learning rate of the AC network. Hence, the network allows the UAV to effectively learn its trajectory on the fly. 
% A suboptimal policy can be obtained through iterations as in \eqref{eqAC}.
A comprehensive workflow of the proposed lifelong RL solution with AC network of the UAV is illustrated in Fig.~\ref{figCom}.

\section{Simulation Results and Analysis}\label{sec:simulation}

% In this section, we perform extensive simulations to evaluate the effectiveness of the proposed algorithm.

\subsection{Simulation Environment and Settings}

% Field: shape, points, distribution, numbers,  data arrival rate

In our simulations, we consider $N=6$ IoT devices distributed within a square region with a side length of \si{1}{km}. 
The horizontal and vertical sides of the region are aligned with the $x$ and $y$ axes, respectively.
We consider the use of Mica2 chips \cite{hill2002mica} on IoT devices. 
As such, the corresponding parameters are adopted such that
$\epsilon_{\textrm{max}} \in [3 \times 10^6,8 \times 10^6]$ cycles/slot, and $\kappa_{i} = 10^{-21} \rm J/cycles^3$. 
Moreover, the packet size  $a_{i,t}$ follows a Gaussian distribution with mean $\bar{a}_{i,t} \in [1 \times 10^7, 5 \times 10^7 ] \, \textrm{cycles}$ and standard deviation $\sigma_{i,t} =  5 \times 10^6 \, \textrm{cycles}$ for each environment. In addition, we consider that the duration of each environment follows a normal distribution with unique parameters for each device ranging from $100$ to $550$ timeslots.
Moreover, each duration of a time slot is considered to be \SI{1}{s}. In addition, we consider $1500$ episodes while having each episode corresponding to $3000$ timeslots. 

% $\bar{a}_{i,t} = 20 \textrm{kb}$

% $\sigma_{i,t} = 4$
% UAV parameters table, coordinate, velocity

The initial location of the UAV is at the origin of the square region, i.e., $\boldsymbol{l}_0=(0,0)$.
We consider the range of the flying velocity of the UAV to be $\SI[per-mode=repeated-symbol]{10}{\m\per\second} \leq  v_m \leq  \SI[per-mode=repeated-symbol]{40}{\m\per\second}$.  The rest of the propulsion energy parameters of the rotary-wing UAV are found in Table \ref{Table:UAV}.

\begin{table}
	\caption{\small Propulsion energy parameters of the UAV.}
	\centering 
	\begin{threeparttable}
	\begin{tabular}{lll}
	%{m{15mm} m{70mm} m{18mm}}
	Parameters  & Simulation Value  & \\
	\midrule\midrule $P_0$  &  $23.661$ &   \\
	\cmidrule(l  r ){1-3}$P_i$ & $88.627$ &  \\ 
	\cmidrule(l r ){1-3}  $v_{\textrm{tip}}$ & $120$ $\textrm{m/s}$ &  \\
	\cmidrule(l r ){1-3}  $v_0$ & $4.03$ &  \\
	\cmidrule(l r ){1-3}  $d_0$ & $0.6$ &  \\
	\cmidrule(l r ){1-3}  $s$ & $0.05$ &  \\
	\cmidrule(l r ){1-3}  $\rho$ & $1.225$ $\textrm{kg/m}^3$  &  \\
	\cmidrule(l r ){1-3}  $A$ & $0.503$ $\textrm{m}^3$ &  \\
	\midrule\midrule
	\end{tabular}
	% \begin{tablenotes}
	% \item[*] Throughout the paper. 
	% \end{tablenotes}
	\end{threeparttable}
	\label{Table:UAV}
\end{table}

% \subsection{Training and Testing Phase}
% Train and test data/procedure:

\begin{figure}
	\centering
	\includegraphics[width=3.8in]{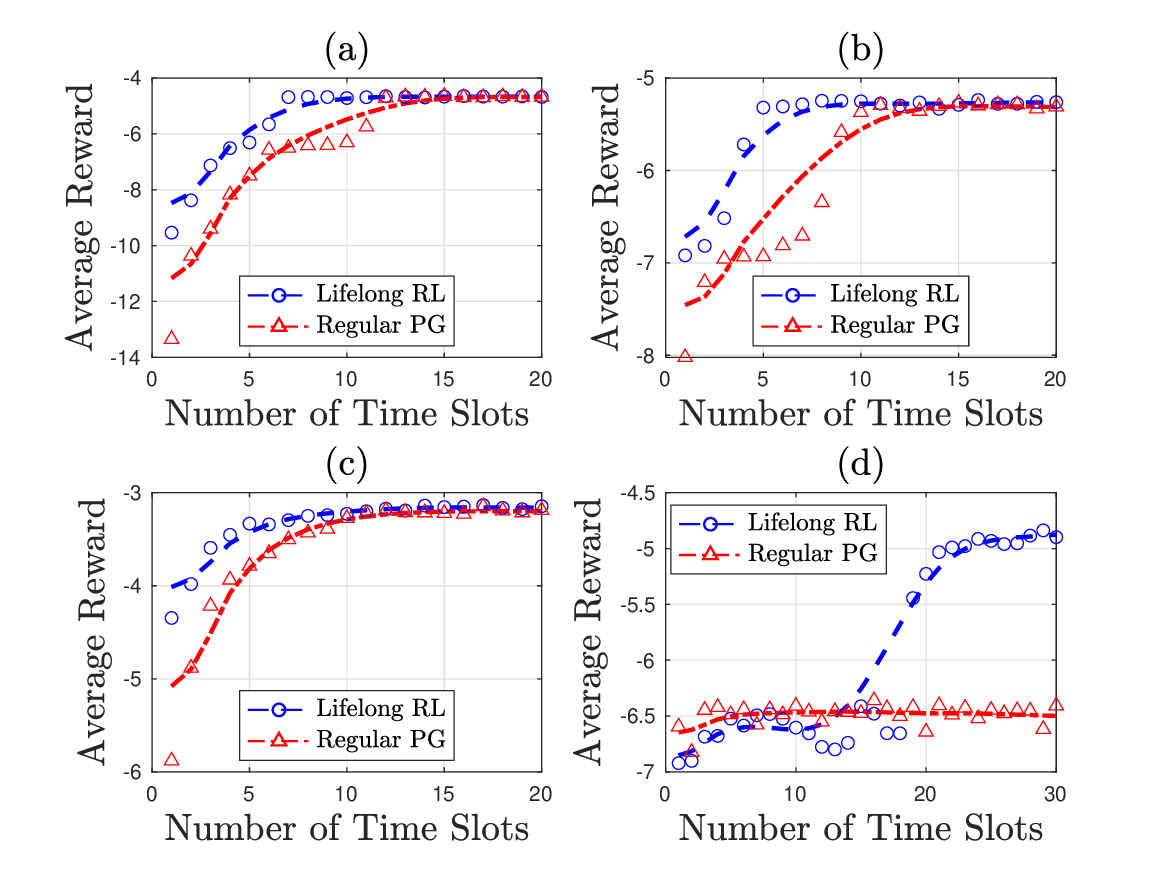}
	\caption{\small Average reward of the regular PG and our proposed zero-shot lifelong learning method for $4$ different examples.}
	\label{fig: single comp}
\end{figure}

\begin{figure}
	\centering
	\includegraphics[width=3.8in]{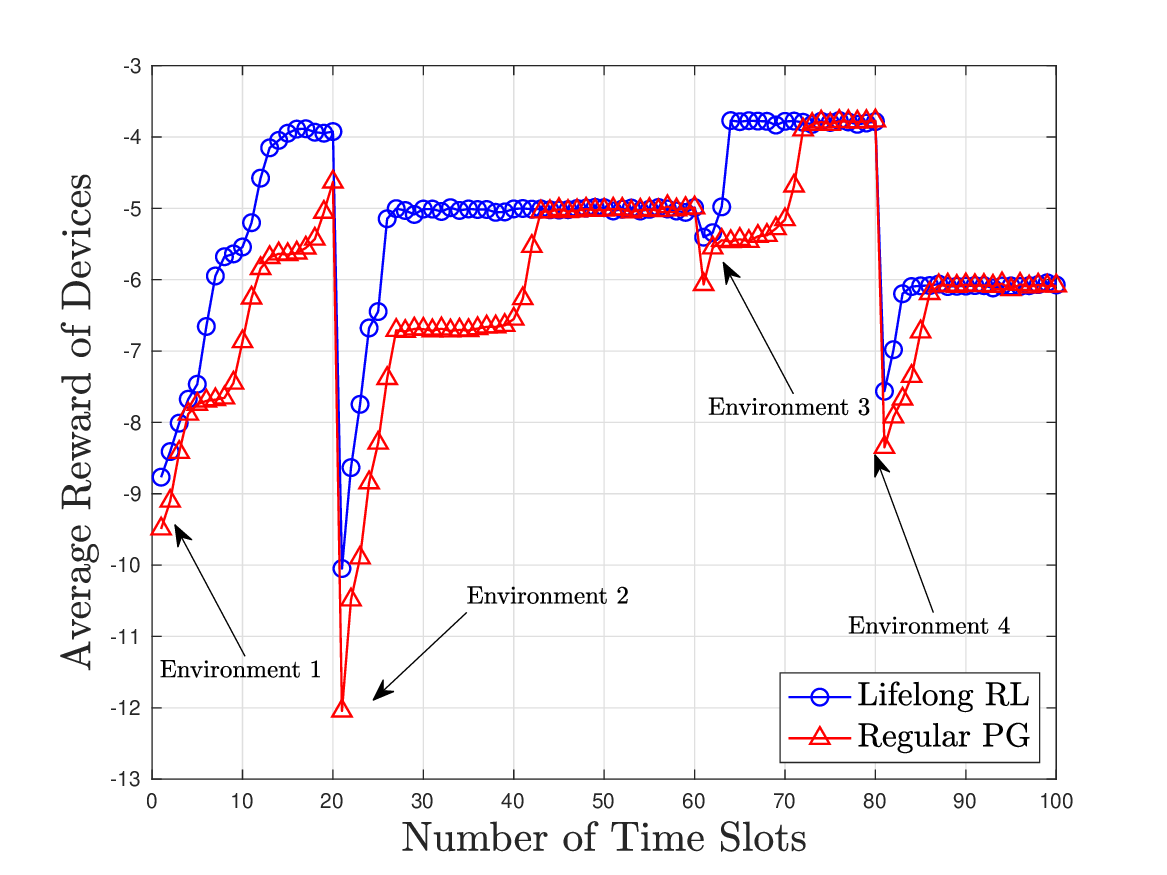}
	\caption{\small Sequential learning for the PG method and zero-shot lifelong learning method over $4$ environments. \vspace{-0.1cm}}
	\label{fig: mul tasks}
\end{figure}

\subsubsection{Training and Testing Procedure}
% Train and test data/procedure:

% In the below, the procedure of training and testing will be presented. 
% For the training procedure, 
% we first learn the optimal policies for a bunch of independent tasks.
% Next, these tasks will be randomly distributed to all the devices according to the setup of the devices.
On the one hand, the training procedure includes the UAV randomly flying between the devices to update the policy of each device sequentially so that it learns the knowledge base and the feature base. 
This process continues until all devices have been visited and all the environments have been experienced. 
The knowledge base and the feature base are then updated and refined with each interaction between the UAV and devices.

On the other hand, the devices and their associated environment are considered to be randomly chosen.
The UAV, equipped with its trained knowledge base and feature base, visits these devices randomly to update their policies using the zero-shot method. 
Following the training of the knowledge base and feature base, the trajectory of the UAV is learned using an AC network. 
% During this stage, 
The policy update method for all devices is the same, while the AC network learns and optimizes the trajectory of the UAV based on the available information and feedback from the devices.
Moreover, the AC network has two hidden layers. The action head applies Softmax to yield the probability of the UAV visiting the devices at its next destination. In addition, a clamp function is used to capture the velocity of the UAV within the designated velocity bounds. 
Additionally, the value head is used to capture the loss in the policy. Herein, we adopt zero grad as the default optimizer.

\subsubsection{Benchmarks and Baselines}

We consider multiple benchmarks to compare with our proposed solution.
% The comparison simulation is divided into two parts: the Lifelong Reinforcement Learning method part and the Actor-Critic part.
Hence, our baseline for lifelong RL solution comprises the regular
%  we use the following methods as a comparison:
\emph{policy gradient (PG) method}. Here, instead of using the zero-shot method to enable a warm-start policy, we consider the regular PG as the base learner for each environment.

Furthermore, we compare our proposed AC network used for optimizing the UAV with the following baselines:
\begin{itemize}
	\item \emph{Random method}: The UAV chooses its flying destination randomly and flies at constant velocity of \SI[per-mode=repeated-symbol]{20}{\m\per\second}.
	\item \emph{Force method}: We determine the visit interval for each device by considering the frequency of change in the device's environments. Similar to the random method, the flying velocity is set to \SI[per-mode=repeated-symbol]{20}{\m\per\second}.
	\item \emph{Value-based method}: We consider a two-layer neural network for Q-learning to acquire the flight trajectories of the UAV. 
\end{itemize}

\subsection{Results and Discusions}

\begin{figure}
	\centering
	\includegraphics[width=\linewidth]{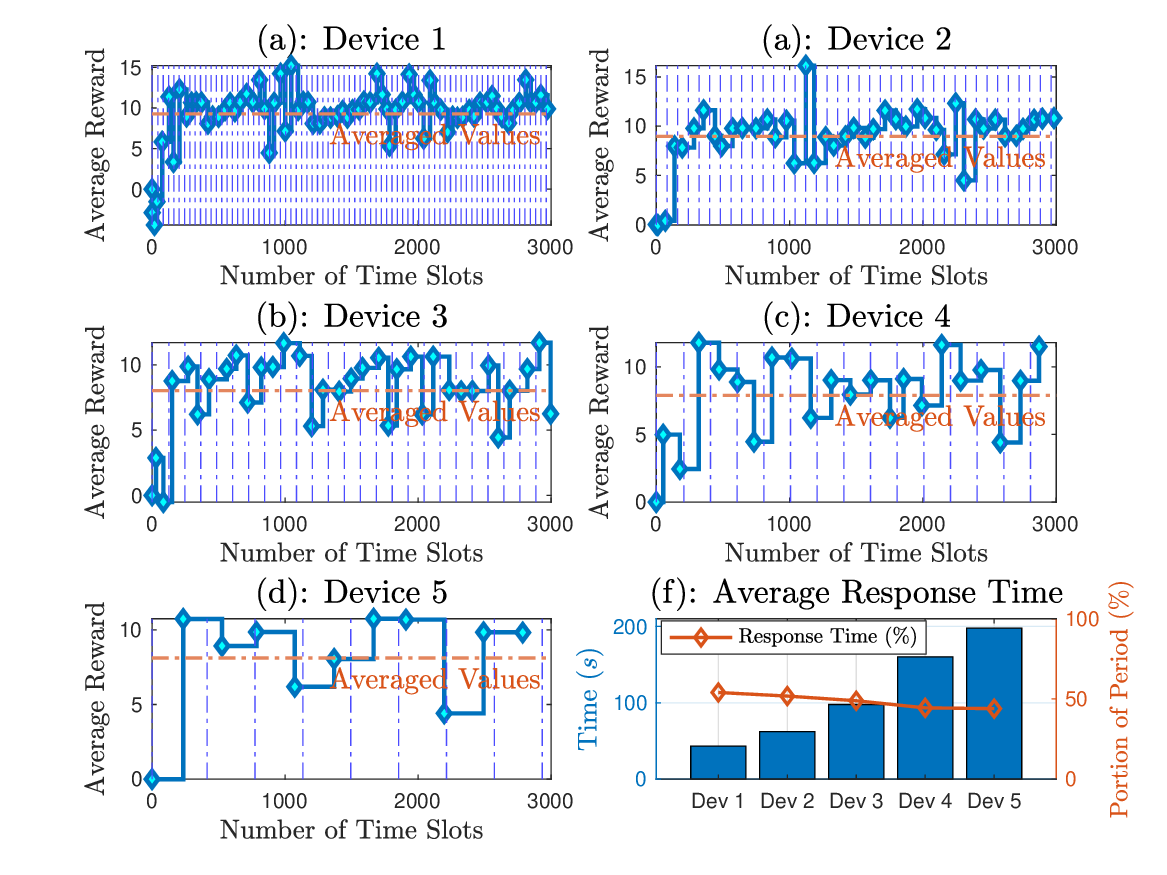}
	\caption{\small History of visits performed by the UAV to $6$ devices having different environmental periods.}
	\label{fig:TimeSlot}
\end{figure}

Fig.~\ref{fig: single comp} compares our proposed zero-shot lifelong learning method with the regular PG method on a set of new environments. From Fig.~\ref{fig: single comp}, we can see that four independent environments are presented as examples to validate our approach. Unlike regular algorithms that need to start with a random initial policy, our method can provide a better warm-start policy due to the knowledge transfer incorporated in zero-shot lifelong learning. 
With this improved starting policy, the convergence time is greatly improved. In particular, our proposed method converges faster than the regular PG method, achieving a $50\%$ improvement in the best case and a $25\%$ improvement in the worst scenario case. In addition, our approach yields a $10\%$ improvement in average reward at the beginning of an environment compared to the random initial policy.
It's important to emphasize that in Fig.~\ref{fig: single comp}(d), our approach attains the global optimum by harnessing accumulated knowledge, a performance significantly superior to that of the baseline algorithm, which merely reaches the local optimum.

\begin{figure}
	\centering
	\includegraphics[width=3.8in]{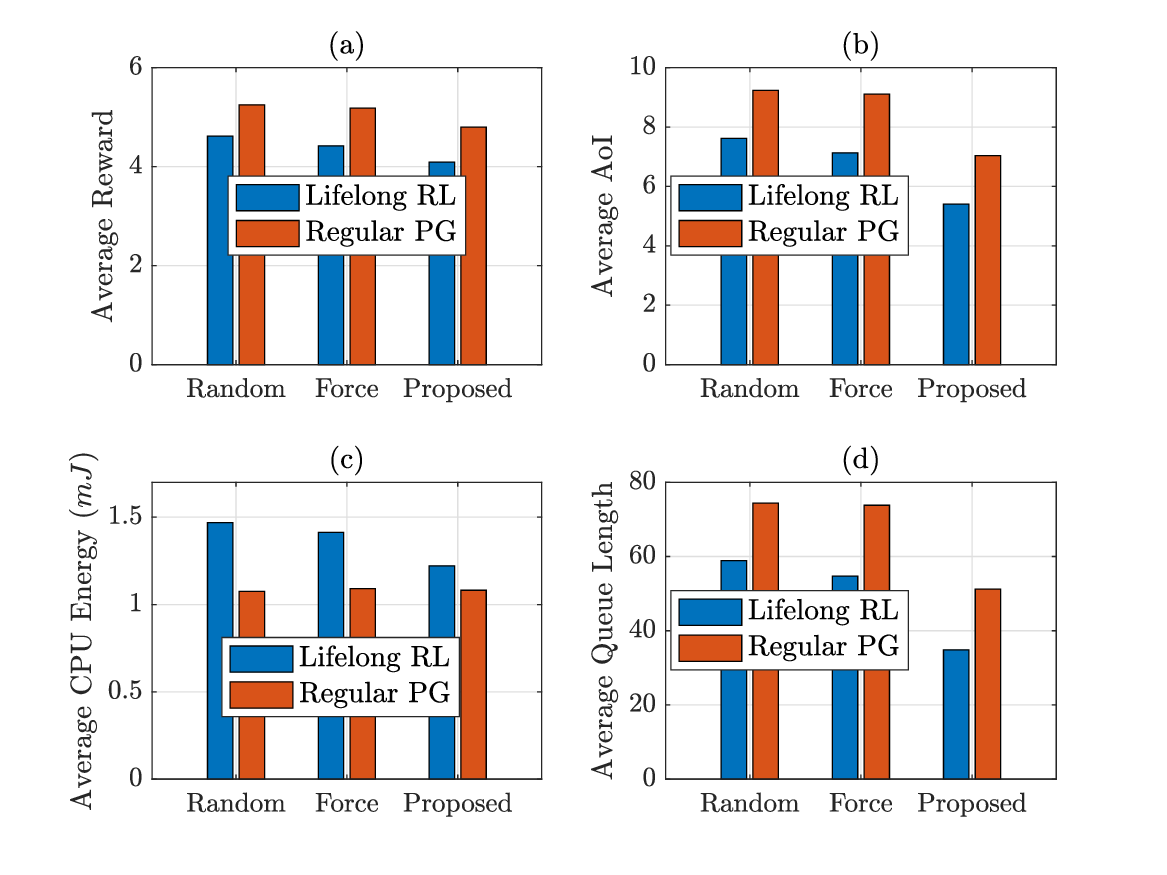}
	\caption{\small Comparison for the average (a) reward, (b)  AoI, (c) consumed CPU energy, and (d) queue length between the proposed lifelong RL method and the regular PG method.}
	\label{fig:PG_LL}
\end{figure}

Fig.~\ref{fig: mul tasks} shows the capability of our proposed lifelong solution in comparison to the PG method under a series of sequential environments encountered by a single IoT device. 
% the learning process of the zero-shot lifelong for multiple environments on a single device.
%As can be seen, the device goes through different environments while the regular PG is used as a comparison. 
We note that the initial policy for both of the aforementioned methods is randomly generated and identical.
% We would like to highlight that the initial policy for both of the aforementioned methods is randomly generated and identical.
From Fig.~\ref{fig: mul tasks}, our proposed algorithm provides better warm-start policy such that the starting reward is $33.7\%$ higher than that achieved by regular PG method. Furthermore, it is also evident that the mean average reward of our proposed algorithm surpasses that of the regular PG method by $8.3\%$. This improvement can be attributed to the superior warm-start policy, which effectively reduces the convergence time.
Clearly, the average reward demonstrates ongoing improvement, even as new environments continuously emerge.
This underscores the sustainability of our lifelong learning algorithm as an evolving algorithm.
% This ensures the sustainability of our lifelong RL solution as a continually evolving algorithm.
% These three environments constitute a non-stationary environments. In addition, our method works well in non-stationary environments with multiple tasks.

% Each device goes through several environments. It can be seen that our method works for non-stationary environments in a continuous way. This is essential in real world practice.

Fig.~\ref{fig:TimeSlot} shows an example of the visits performed by the UAV to IoT devices with different environmental periods
\footnote{For simplicity, we now utilize the absolute value of the reward function from here on. Hence, the rewards in upcoming figures will be represented as positive values.}. 
This example includes $5$ devices where the duration of each environment is constant for each device. 
Here, the duration of each environment increases incrementally from 40 to 440 time slots as we as we transition from device $1$ to device $5$, respectively.
%, having the time between two consecutive slotted lines correspond to one environment period.
%We also show red horizontal line is the averaged reward value throughout the periods. 
From Fig.~\ref{fig:TimeSlot}, it is evident that the UAV visits the IoT devices successively at different rates. In fact, as the duration of the environment increases, the likelihood that the UAV visits the corresponding IoT device decreases.
In particular, $39\%$ of the UAV's visits are allocated to device $1$ while only $5\%$ of the its visits are designated for device $5$. 
% Thus, this distribution is directly proportional. 
This distribution aligns with the frequency of changes in the environment of each IoT device.
As such, the averaged value of the reward for device $5$ is $6.7212$ which is remarkably better than that of device $1$ having a value of $9.2810$.
This is due to the frequent changes in the environment that severely degrade the rewards and necessitate additional visits by the UAV to compensate for frequent policy updates.
% From Fig.~\ref{fig:TimeSlot} (f), it is evident that the UAV exhibits precise control over its timing for visits to the devices.
From Fig.~\ref{fig:TimeSlot}(f), we can observe that the UAV responds to the environmental changes within a consistent time interval percentage across all  devices. This response time accounts for $54.03\%$ of the period for device $1$, which slightly decreases to $44.44\%$ for device $5$.
It is evident that the UAV exhibits accurate control over its timing for visits to the devices with respect to each environment.
% However, it is worthwhile noting that the energy constraints of the UAV may introduce a delay gap in the immediate response to environmental changes. Clearly, this relates to balancing the cost of AoI and energy consumption.
% Consequently, the proportion of response time within each environmental period decreases from $54.03\%$ for device $1$ to $44.44\%$ for device $5$.}
This is pertained by the proposed AC network, which empowers the UAV to effectively balance the disparities introduced by varying environmental periods.

Fig.~\ref{fig:PG_LL} provides a thorough comparison between our lifelong RL method and the regular PG method for different UAV flying strategies. 
This comparison is carried out on multiple fronts that include the average reward, AoI, and CPU energy consumption, and queue length. On the one hand, our proposed AC method for the UAV outperforms the random and force methods on all of the aforementioned levels. On the other hand, our lifelong RL solution achieves a remarkable upperhand in comparison to the regular PG method. 
In particular, from Fig.~\ref{fig:PG_LL}(a), the average reward of the lifelong RL is $7.69\%$
% 实际的是 $11.69\%$， 为了不比fig8的结论高，所以选了个小的。
lower than regular PG. This is attributed to the ability of the zero-shot method to facilitate a warm-start policy for each new environment, whereby the the duration of suboptimal policies is effectively reduced.
%In addition, the average reward of the proposed method is better than Random and Force both in respect of lifelong RL and regular PG. This is because the proposed strategy is able to take the devices' reward into consideration. 
In addition, Fig.~\ref{fig:PG_LL}(b) shows that the proposed lifelong RL solution acheives a $21.42\%$ reduction in AoI in comaprison to the regular PG method. In addition, Fig.~\ref{fig:PG_LL}(d) shows that  the lifelong RL method can achieve a reduction of $47.04\%$ in queue length compared to regular PG. This gain arises is due to the efficient utilization of the allocated CPU resources to process the arrived data packets at each IoT device. 
%it is evident that the utilization of the proposed method results in the lowest AoI compared to other strategies. To be specific, the AoI of proposed method exhibits a reduction of $24.23\%$ compared to the Force method and a $29.08\%$ decrease compared to the Random method.
Unlike the trends of the reward, AoI, and queue length, Fig.~\ref{fig:PG_LL}(c) showcases that the regular PG consumes less CPU energy compared to lifelong RL solution. Nevertheless, this comes at the expense of employing suboptimal interacting policies, which in turn fail to strike a balance between AoI and CPU energy consumption. This subsequently leads to increased values of rewards, AoI, and queue length. 
% which inappropriately allocate more CPU energy, ultimately leading to elevated AoI.
% {\color{red}Also, the Force method has the highest CPU energy consumption, such that LLRL is $44.24\%$ higher than regular PG in Force method. This is because that LLRL here is trying to get lower AoI under bad flying strategies while more energy is used.}
% Fig.~\ref{fig:PG_LL} (d) shows the average queue length of IoT devices. The proposed method can achieve a $47.04\%$ reduction in queue length compared to regular PG when applied in conjunction with the AC method. This gain arises from the ability of our proposed method to efficiently utilize the allocated CPU resource to process the arrived data packets at each device. 
From Figs.~\ref{fig:PG_LL}(a) to ~\ref{fig:PG_LL}(d), it is evident that our proposed method outperforms the PG method in terms of the total reward of devices, AoI and queue length. However, this improvement comes at the cost of increased CPU energy consumption due to the inherent trade-off between AoI and CPU energy consumption.

% \begin{figure}[htbp]
% 	\centering
% 	\includegraphics[width=3.8in]{Fig_three_bar.eps}
% 	\caption{The comparison between different mu.}
% 	\label{fig:MuBar}
% \end{figure}

\begin{figure}
	%\centering
 \flushleft
	\includegraphics[width=\linewidth]{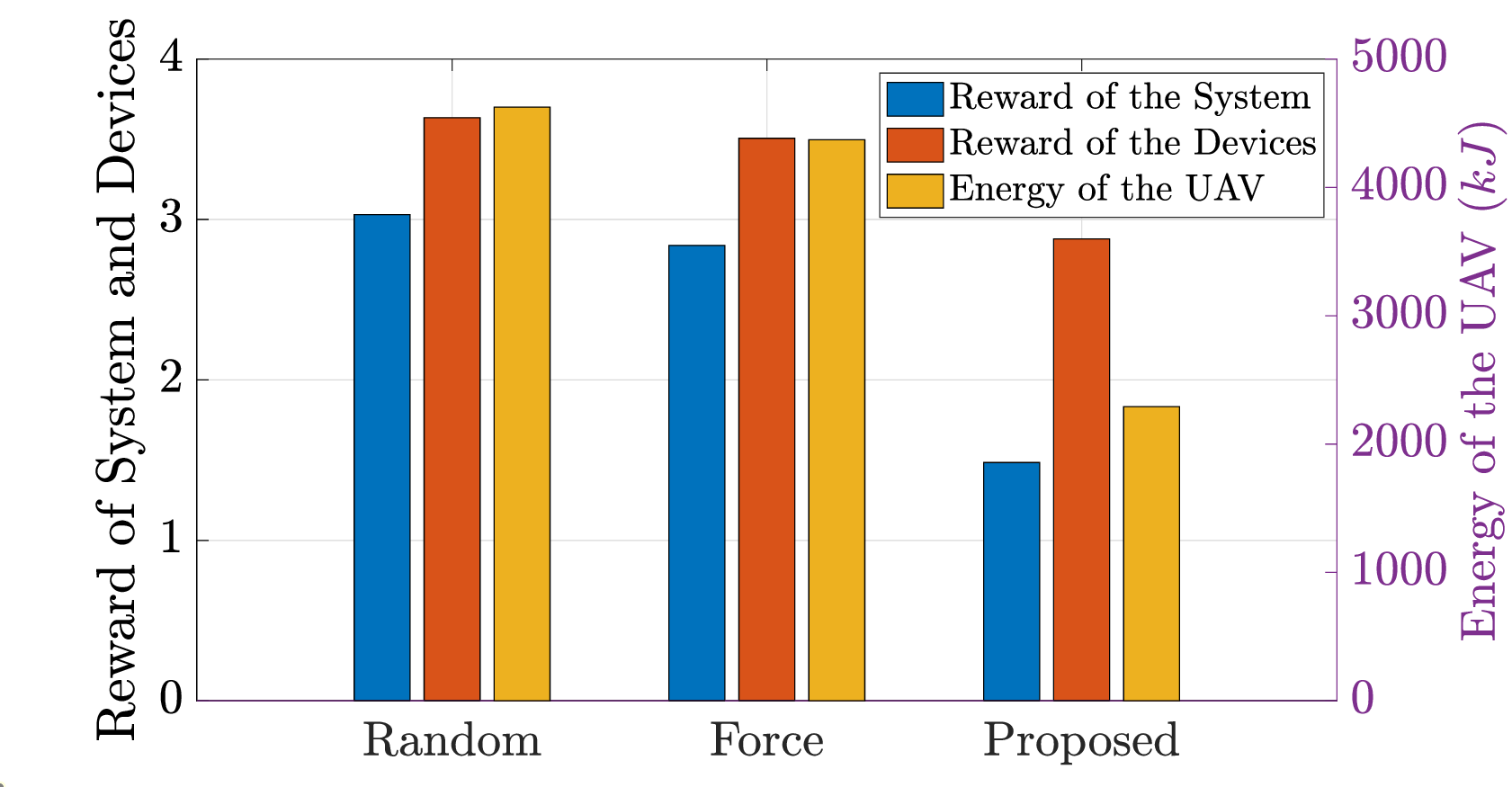}
	\caption{\small The rewards and energy consumption of the random, force, and proposed AC methods for the UAV flying strategies.}
	\label{fig:Costbar}
\end{figure}

Fig.~\ref{fig:Costbar} showcases how the proposed AC method outperforms the random and force methods in terms on the rewards and energy efficiency for the UAV. 
From Fig.~\ref{fig:Costbar}, our proposed method attains around $49\%$ improvement in terms of system rewards with respect to the other baselines.
%superior system reward, amounting to a $47.75\%$ reduction compared to the Force method and a $50.98\%$ decrease compared to the Random method. 
One of the main reasons for this is the precise selection of destination and velocity that results in a significant reduction in the UAV's energy consumption that reaches $2292$ kJ. This corresponds to a substantial energy savings of $48.5\%$ as compared to the force method and random methods.
From a different perspective, the rewards of the devices that are incorporated in the trade-off controlled by $\mu$ play a significant role in enhancing the overall reward of the system.
%  thanks to insights derived from the AC model. 
% The UAV's selection is able to empower the LLRL with the careful flying decisions.
Indeed, the AC method empowers the lifelong RL by equipping it with prudent flying decisions, thereby mitigating losses pertaining to suboptimal interacting policies and enhancing the rewards of the devices compared to the other two methods.
These results verifies that the AC method can learn the varying distribution of environmental periods across multiple devices.
% This is due to the fact that, our method learns the varying durations of the environments for each device and optimizes the UAV's flights accordingly. 
Meanwhile, the force method effectively mitigates the decrease in rewards of the devices caused by suboptimal policies upon comparison to the random method.
% The Force method is able to catch the key changes of the environments. As such, it can reduce the loss of reward due to policy expiration compared to random strategies. 
However, due to its constant velocity and lack of energy consideration, the overall reward of the force method remains inferior to our proposed approach.
% Even though the Force method chooses the flying destination without taking the flying velocity into consideration, the flying energy is its advantage because 

% With respect to the energy consumption of the UAV, our proposed method greatly reduces the flying energy. This improvement stems from the utilization of the AC method, which takes the UAV's flying energy into account through meticulous selection of destination and velocity.
% At the same time, the cost of the system is slightly decreased. This is the averaged cost of the devices with the trade-off factor $\alpha$ of 0.5.

\begin{figure}
	\centering
	\includegraphics[width=\linewidth]{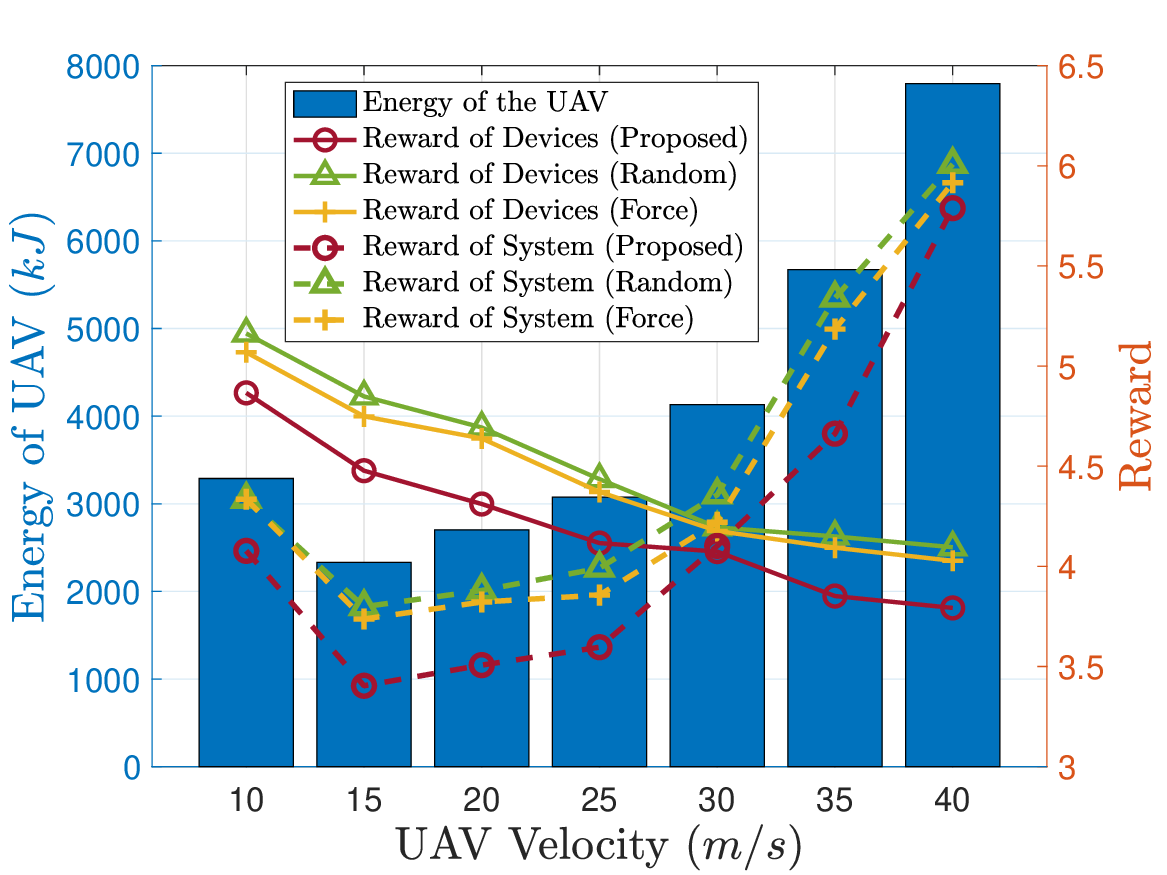}
	\caption{\small Reward of system and devices along with energy consumption of the UAV as a function of the velocity of the UAV.}
	\label{fig:PG_Velocity_R_E}
\end{figure}

Fig.~\ref{fig:PG_Velocity_R_E} shows the system performance in terms of system and device rewards across different UAV flying methods for different velocities of the UAV. From Fig.~\ref{fig:PG_Velocity_R_E}, the energy consumption of the UAV decreases at first, after which it increases with the incermental increase of the velocity from \SI[per-mode=repeated-symbol]{10}{\m\per\second} to \SI[per-mode=repeated-symbol]{40}{\m\per\second}. This is consistent with the energy consumption model of the rotary wing UAV \cite{zeng2019energy}.
%  where the lowest UAV flying energy appears around \SI[per-mode=repeated-symbol]{15}{\m\per\second}.
In addition, the reward of the devices decreases $22.06\%$ as the velocity increases from \SI[per-mode=repeated-symbol]{10}{\m\per\second} to \SI[per-mode=repeated-symbol]{40}{\m\per\second}.
Here, maintaining a higher UAV velocity enables more frequent device visits, which can facilitate the development of improved interacting policies. 
% These up-to-date policies allow the devices to adapt to the non-stationary environments more effectively. 
Meanwhile, the force and random methods have a similar performance. In fact, the reward of the devices in the random and force method are on average $5.78\%$ and $4.61\%$ higher than our proposed AC method, respectively.
Clearly, it is evident from Fig.~\ref{fig:PG_Velocity_R_E} that the system's reward is proportional to the energy consumption of the UAV. This is reflected in a remarkable $70.05\%$ difference between the highest and lowest rewards. Clearly, the UAV's flying energy significantly impacts the overall reward of the system.  
% Likewise, the AC method outperforms Random and Force method on system reward because of its capability to ptomptly respond to environmental changes. 
% Additionally, our proposed approach outperforms the Random and Force methods. This superiority is attributed to our method's ability to promptly respond to environmental dynamics.
This underscores the significance of our proposed method in learning and optimizing the flying strategy of the UAV.

\begin{figure}
	\centering
	\includegraphics[width=\linewidth]{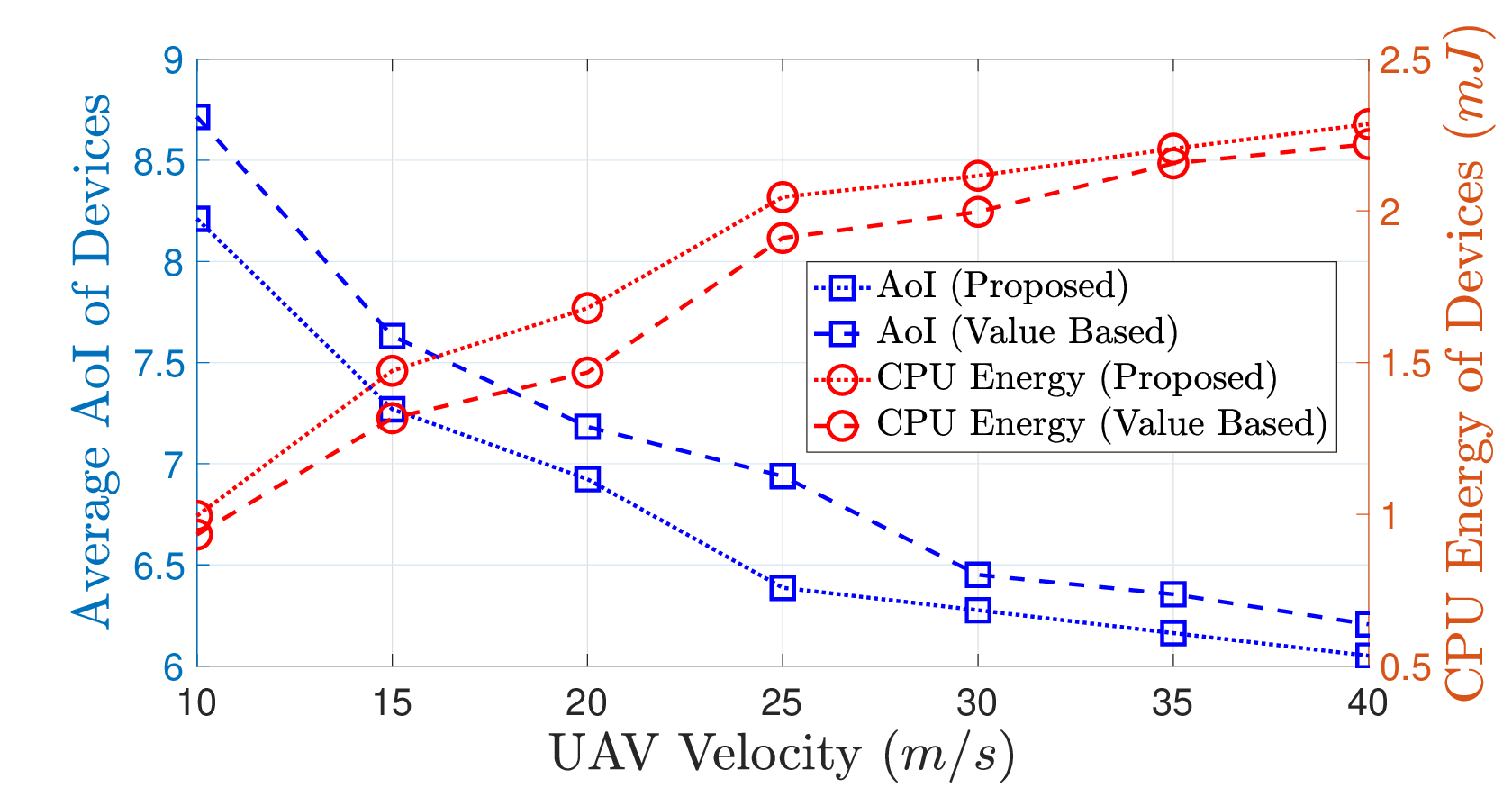}
	\caption{\small The AoI and CPU energy consumption as a function of the velocity of the UAV for the proposed and value-based methods.}
	\label{fig:PG_Velocity_AoI_CPU}
\end{figure}

\begin{figure}
	\centering
	\includegraphics[width=3.7in]{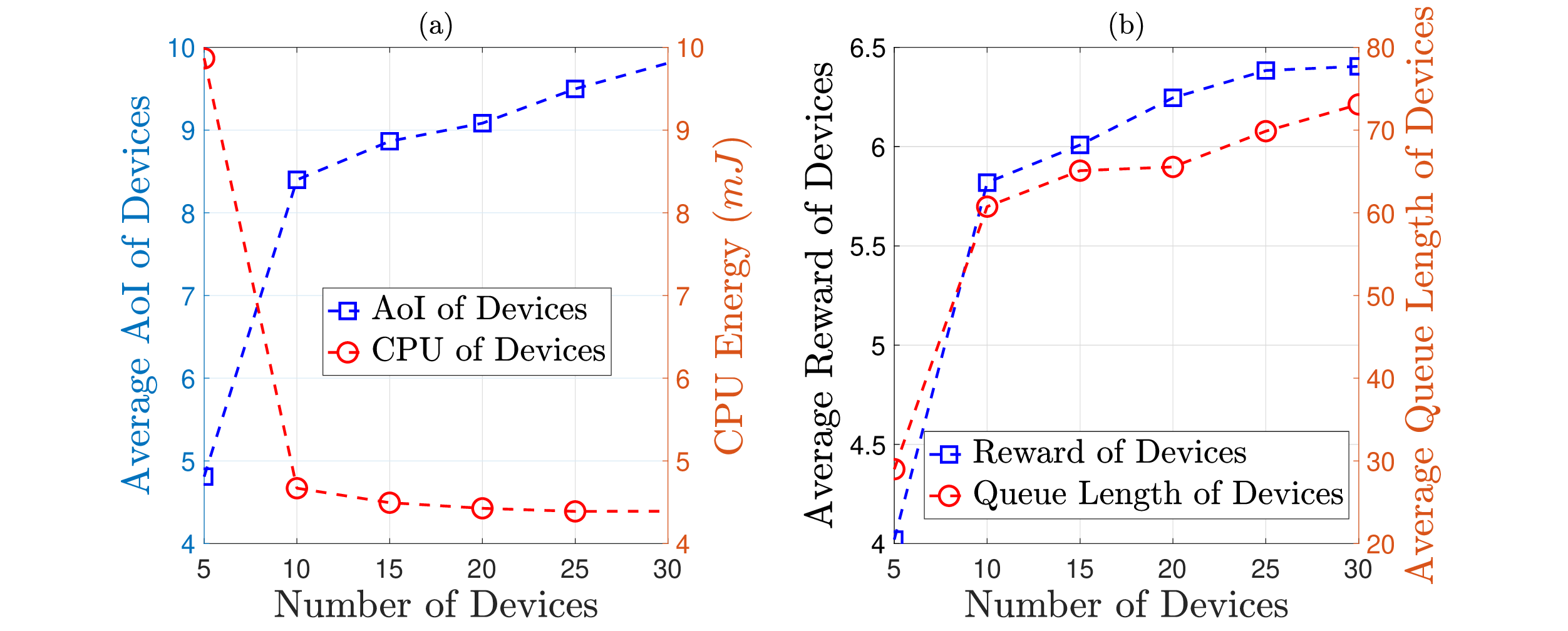}
	\caption{\small The (a) AoI and CPU energy consumption, and (b) reward of IoT devices and queue length as a function of the number of devices.\vspace{-0.4cm}}
	\label{fig:PG_NumDev}
\end{figure}

Fig.~\ref{fig:PG_Velocity_AoI_CPU} illustrates the relationship between the AoI and CPU energy consumption of devices for different UAV velocities and for two methods: a) the proposed AC method and the b) the value-based method.
%as UAV's velocity increases from \SI[per-mode=repeated-symbol]{10}{\m\per\second} to \SI[per-mode=repeated-symbol]{40}{\m\per\second}.
From Fig.~\ref{fig:PG_Velocity_AoI_CPU}, the AoI in the proposed method decreases by $26.30\%$ 
%from $8.21$ to $6.05$
as the velocity increases from \SI[per-mode=repeated-symbol]{10}{\m\per\second} to \SI[per-mode=repeated-symbol]{40}{\m\per\second}. 
This is due to the fact that allowing the UAV to conduct frequent visits enables the devices to improve their interacting policies through lifelong RL.
% This is attributed to the devices acquiring improved interaction policies due to the more frequent visits by the UAV.
Accordingly, this results in an increased energy demand to process incoming packets at the devices. As such, the energy consumption increases from $0.99$ mJ to $2.29$ mJ to account for the additional CPU processing of the devices. Subsequently, this decreases the AoI and shortens the queue lengths. 
% In contrast, the reward of devices decreases as the UAV velocity increases.
% Nevertheless, by leveraging the trade-off factor $\mu$, the overall reward of the devices encompassing both AoI and CPU energy consumption decreases.
%Furthermore, we have conducted a comparative analysis between our proposed method and a value-based learning method, specifically Q-Learning, with a focus on UAV's flight strategies.
% Hence, it is evident that our proposed method outperforms the value-based method by achieving a $9.84\%$ improvement in terms of the rewards of the devices.
This decrease is a direct implication of utilizing additional CPU energy to reduce queue length, and simultaneously, enhancing the data freshness.

Fig.~\ref{fig:PG_NumDev} shows the performance of the devices as the number of the IoT devices varies from $5$ to $30$. The performance is comprising of their averaged AoI and CPU energy consumption, on one hand, and the reward and queue length, on the other hand. From 
In fact, as the number of devices increases, the UAV can perform a lower number of visits for each device.  
Hence, suboptimal interacting policies that degrade the performance of each device arise accordingly. Thus, from Fig.~\ref{fig:PG_NumDev}(a), we can observe that the AoI increases from $4.81$ to $9.81$, as the number of devices increases. 
% This illustrates the capacity of the UAV in IoT network, can further guide the study of number of UAVs. 
This observation underscores the importance of constraining the UAV's flying range to achieve optimal efficiency.
In contrast, the CPU energy consumption experiences a significant decrease that reaches up to $55\%$ as the number of devices increases. Here, the energy consumption reaches a plateau when the number of devices exceeds 10. 
Evidently, poor interacting policies can bring drastic effects to the energy efficiency of the CPU. 
% With the increase of the number of devices, the AoI performance improves along with the total cost of the devices. 
% This is due to the fact that the UAV allocates fewer visits to each individual device as the number of devices increases. 
% Nevertheless, the CPU energy consumption of the devices decreases, which is reasonable since an ineffective data packet processing policy could lead to an inappropriate CPU allocation.
From Fig.~\ref{fig:PG_NumDev}(b),  the reward of the devices increases by $59.29\%$ as the number of devices increases. Furthermore, the average queue length for all the devices also experiences an increase due to the delayed processing of incoming data packets. Hence, as the number of devices increases, we can see that the having many devices will inevitably lead to an increase in the rewards. However, this comes at the expense of delayed UAV response and an elevated queue length.

\section{Conclusion}\label{sec:conclusion}

In this paper, we have proposed a novel UAV-aided lifelong RL solution that leverages a UAV to adapt the policies of IoT devices in non-stationary environments.
% provide fast adaptation for IoT devices in non-stationary environments. 
Our proposed method aims to continuously optimize the data freshness and energy efficiency of IoT devices, while efficiently utilizing the energy resources of the UAV. Hence, we have designed a lifelong RL solution that leverages a shared knowledge base and feature base to acquire the optimal policies for the dynamic non-stationary environments.
In addition, we have proposed a zero-shot method to determine the warm-start policies for unseen environments. 
To efficiently utilize the energy resources of the UAV as it learns from the environments, its corresponding flying trajectory and velocities are optimized by adopting an AC network solution mechanism. 
% The proposed framework also enables the rapid adaptation by determining the best policies for unseen environments with zero-shot method utilized. 
%Ultimately, our proposed method offers a versatile and adaptive solution with potential to real-world applications.
Our simulations have validated that the proposed lifelong approach yields significant performance gains in terms of minimized AoI and optimal energy efficiency for both IoT devices and the UAV, respectively.

\ifCLASSOPTIONcaptionsoff
  \newpage
\fi



%\bibliographystyle{IEEEtran}
%\begin{thebibliography}{1}
%\end{thebibliography}
\bibliographystyle{IEEEtran}
\bibliography{bare_jrnl_comsoc}

% \end{IEEEbiography}


%\newpage



%\vfill


\end{document}